\documentclass[12pt]{article}
\usepackage[utf8]{inputenc}
\usepackage{comment}
\usepackage{cite}
\usepackage[absolute]{textpos}
\input epsf.sty
\usepackage{xcolor}
\usepackage{verbatim}         	
\usepackage{graphicx}				% Use pdf, png, jpg, or eps§ with pdflatex; use eps in DVI mode
\usepackage{hyperref}
\usepackage{amssymb}
\usepackage{amsmath}
\usepackage{xcolor}
\newcommand{\be}{\begin{eqnarray}\displaystyle}
\newcommand{\ee}{\end{eqnarray}}
\newcommand{\nn}{\nonumber}

\newcommand{\h}{\bar{h}}

\newcommand{\p}{\partial}

\hypersetup{
  colorlinks   = true, %Colours links instead of ugly boxes
  urlcolor     = blue, %Colour for external hyperlinks
  linkcolor    = blue, %Colour of internal links
  citecolor   = red  %Colour of citations
}

\title{Ward identity for loop level soft photon theorem for massless QED coupled to gravity.}

\author{}
\date{}					
\begin{document}	
\begin{textblock}{5}(6,1)
 \color{red}\Large $||$ Sri Sainath $||$
\end{textblock}
\color{black}
\maketitle
\centerline{\large {  Sayali Atul Bhatkar,  }}

\vspace*{4.0ex}

\centerline{\large \it Indian Institute of Science Education and Research,}
\centerline{\large \it  Homi Bhabha Rd, Pashan, Pune 411 008, India.}

\vspace*{1.0ex}
\centerline{\small E-mail: sayali014@gmail.com.}
\vspace{3cm}
\textbf{Abstract}\\

% We study the effect of long range forces on asymptotic dynamics of massless scalars. 
Strominger and his collaborators pioneered the study of equivalence between soft theorems and asymptotic conservation laws. We  study this equivalence in the context of loop level subleading soft photon theorem for massless scalar QED in presence of dynamical gravity. Motivated by Campiglia and Laddha \cite{1903.09133}, we show that the Sahoo-Sen soft photon theorem \cite{1808.03288} for loop amplitudes is equivalent to an asymptotic conservation law. This asymptotic charge is directly related to the dressing of fields due to long range forces exclusively present in four spacetime dimensions. In presence of gravity, the new feature is that soft photons also acquire a dressing due to long range gravitational force and this dressing contributes to the asymptotic charge. %We finally show the corresponding Ward identity is equivalent to the Sahoo-Sen soft photon theorem.
% (both )
\newpage

\tableofcontents
\vspace{2.5cm}

\section{Introduction and Result}
%Asymptotic symmetries refer to the subgroup of the gauge group that acts non-trivially on the physical data.
Asymptotic symmetries strongly constrain low energy physics of gauge theories \cite{1308.0589, 1312.2229, g, qed1,qed2,qed3}. Leading soft theorems are manifestations of asymptotic symmetries. Soft theorems are statements about universal properties of amplitudes in the limit when energy of some of the interacting massless particles is taken to be small \cite{Gell-Mann,Low,weinberg,jackiw}\footnote{Interested readers can look up the references of \cite{1808.03288} and \cite{1703.05448} for recent literature on Soft theorems.}. The equivalence between
the two was first demonstrated in the seminal paper \cite{g}. Similar analysis was carried out for QED in \cite{qed1,qed2,qed3}; it was shown that leading soft photon theorem is equivalent to the Ward identity of the so called large gauge transformations. Large gauge transformations constitute an infinite dimensional subgroup of U(1) gauge transformations.
% The classical version of above problem was studied in \cite{1703.07884}; the authors derived a conservation law in the classical theory itself. The leading soft photon theorem can be understood as quantum version of this conservation equation. The 'diagonal' subgroup is defined by identifying the future gauge parameters with past gauge parameters via anitpodal map.   (diagonal) 

Analogous investigations have been carried out to understand the possible symmetry origins of tree level subleading soft theorem. Ward identity corresponding to Low's subleading photon theorem has been studied in \cite{1605.09677,1407.3814,1605.09731}. The symmetry underlying this Ward identity or its relation to U(1) gauge group is not clear. In \cite{1810.04619}, the authors proved an infinite hierarchy of asymptotic conservation laws for classical electromagnetism and showed that quantum version of the first of these laws is equivalent to Low's subleading soft photon theorem. The authors also provide evidence that suggests that this entire hierarchy is equivalent to the infinite hierarchy of tree level soft theorems proved in \cite{Hamada Shiu, soft inf}. Thus, tree level subleading soft theorems in QED can be related to asymptotic conservation laws though the question of existence of a well defined underlying symmetry still persists. %The story is similar for gravity beyond subleading order : Ward identities for tree level graviton soft theorems beyond subleading order were studied in \cite{1608.00685,1605.09094} and the authors had pointed out similar questions about the corresponding symmetry. % Above statements are true only for tree level physics. 

In this paper we are interested in studying the equivalence between soft theorems and asymptotic conservation laws in presence of loop corrections. The leading soft theorem is true to all loop orders and hence is an exact quantum statement. The Ward identity corresponding to large gauge transformations is also exact. Beyond the leading order, soft theorems receive non-trivial loop corrections in four spacetime dimensions as shown in \cite{1405.1015,1405.1410,1405.3413}. A part of these loop corrections are divergent. In \cite{1701.00496}, the authors showed that these divergent terms can be absorbed by renormalising tree level Ward identity. In the seminal paper \cite{1808.03288}, the authors extended the regulating technique introduced in\cite{yennie} and used it to show that loop effects lead to a new logarithmic soft theorem in four spacetime dimensions. Thus, the subleading soft theorem for loop amplitudes is very different from the tree level subleading soft theorem. This soft theorem is 1-loop exact.

%discussed the new logarithmic modes that appear in the asymptotic region as a result of long range electromagnetic force acting on massive scalars and showed that these logarithmic modes
A natural question arises at this point : is this soft theorem related to a new asymptotic symmetry? The first step in this direction was taken in \cite{1903.09133}. The authors have provided evidence to show that the Sahoo-Sen soft theorem for massive scalar QED has an underlying conservation law. This is quite a remarkable result given the fact that the loop level soft factor has a very complicated structure\cite{1808.03288}. The authors also established a correspondence between the loop level soft factor and the Fadeev-Kulish dressing of massive particles\cite{FK}. It must however be noted that the nature or existence of a well defined symmetry associated to this conservation law is not clear at this point.

In this paper, our aim is to show that the Sahoo-Sen soft photon theorem is equivalent to the asymptotic conservation law given in \eqref{cons} for massless scalar QED in presence of dynamical gravity. Let us first quote the Sahoo-Sen soft photon theorem in presence of gravitational couplings and massless complex scalars \cite{1808.03288} :
\begin{align}
\mathcal{M}_{n+1}(p_i,k)&=\frac{{S}_0}{\omega}\ \mathcal{M}_n(p_i)\  + \ S_{\log}\log{\omega}\ \mathcal{M}_n(p_i)\ + ...\ ,\nn
\end{align}
here, $S_0=\sum_i e_i\frac{\epsilon.p_i}{ p_i.q}$ is the leading soft factor and
\begin{align}
S_{\log} 
 &=\ \frac{ig}{4\pi}\sum_{\substack{i,j; i\neq j \\ \eta_i\eta_j =1}} e_i\frac{\epsilon_\mu q_\rho}{ p_i.q}(p_j^\rho p_i^\mu-p_i^\rho p_j^\mu)-\frac{ig}{4\pi} \sum_i e_i\frac{\epsilon.p_i}{p_i.q}\ \sum_{j,\eta_j=1}q.p_j\ \ \nn\\
& -\frac{1}{4\pi^2}\sum_{i,j;i\neq j}e_i\frac{\epsilon_\mu q_\rho}{ p_i.q}\  \Big[ \frac{e_ie_j}{p_i.p_j}(p_j^\rho p_i^\mu-p_i^\rho p_j^\mu)+g(p_j^\rho p_i^\mu-p_i^\rho p_j^\mu)\log[p_i.p_j]\ \Big]\footnotemark\nn\\
& +\ \frac{g}{4\pi^2}\sum_i e_i\frac{\epsilon.p_i}{ p_i.q}\sum_{j}q.p_j\ \log\ {{p_j.q}}.\label{softthm2}
\end{align}
\footnotetext{Similar to the first electromagnetic term in this line, there could be a potential gravitational term : $$ -\frac{1}{4\pi^2}\sum_{i,j;i\neq j}ge_i\frac{\epsilon_\mu q_\rho}{ p_i.q}\ (p_j^\rho p_i^\mu-p_i^\rho p_j^\mu)$$ that vanishes because of momentum conservation. } 
In above expression, $\epsilon$ is the polarisation vector for the soft photon and $k=\omega q$ is the soft momentum. The indices $i,j$ take values from 1 to $n$, where $n$ is the number of hard particles. $e_i, p_i$ denote the electric charge and momentum of $i^{th}$ hard particle respectively. The momenta and charges are defined including $\eta$ factors such that $\eta_i=1 (-1)$ for outgoing (incoming) particles.  In above expression, we have introduced $g$ to keep track of gravitational terms. We later set $g=8\pi G$=1. An important point to note is that the presence of gravitational coupling modifies the soft photon theorem significantly.

In this paper, we show that above soft theorem is equivalent to the asymptotic conservation law given in \eqref{cons1} :
\be
Q_{\text{1-loop}}^+[V^A_+]\ |_{\mathcal{I}^+_-}\ \  = \ \ Q_{\text{1-loop}}^-[V^A_-]\ | _{\mathcal{I}^-_+}.
\ee 
This charge is very closely related to the long range forces present in four spacetime dimensions. Long range forces affect asymptotic dynamics of various fields nontrivially and lead to a dressing of free fields. $Q_{\text{1-loop}}$ gets contribution from dressing of free fields due to long range forces. Let us note some interesting features of this charge. 
\begin{itemize}
\item
 The leading order dressing of massless scalar field is given by \eqref{phidress}:
\be \phi(x)= -\frac{ie^{ieA^1_r(\hat{x})\log r}}{8\pi^2r}\int d\omega\ [b(\omega,\hat{x})\ e^{-i\omega u}e^{i\omega\log r \frac{h^1_{rr}(\hat{x})}{2}\ } - d^\dagger(\omega,\hat{x})\ e^{i\omega u}e^{-i\omega\log r \frac{h^1_{rr}(\hat{x})}{2}\ }].\nn \ee
$A_r^1$ defined in \eqref{Afalloff} is the electromagnetic dressing and  $h_{rr}^1$ defined in \eqref{h} is the gravitational dressing. The 1-loop charge receives contribution from both electromagnetic and gravitational dressing of massless scalar field. This contribution (given in \eqref{A1u}) can be schematically written as $Q_{\text{1-loop}}\sim \hat{S}_1\ [h_{rr}^1+A_r^1]$, where $\hat{S}_1$ closely resembles the tree level subleading soft operator. %Though the precise structure of dressing of massless scalars is different from that of massive scalars\cite{1903.09133}, the contribution to the charge has very similar structure. =\sum_ie_i\frac{\epsilon^\mu q^\nu}{p_i.q}[p_{i\mu}\p_{i\nu}-p_{i\nu}\p_{i\mu}]
\item
Photons also acquire gravitational dressing \eqref{Adress}:
\be A_\sigma(x)= -\frac{i}{8\pi^2r}\int d\omega\ [a_\sigma(\omega,\hat{x})\ e^{-i\omega u}e^{i\omega\log (r\omega) \frac{h^1_{rr}(\hat{x})}{2}\ } -a_\sigma^\dagger(\omega,\hat{x})\ e^{i\omega u}e^{-i\omega\log (r\omega) \frac{h^1_{rr}(\hat{x})}{2}\ }].\nn\ee 
The leading order gravitational dressing factor i.e. the $\log r$ term is similar for both photon field and massless scalar field. The photon field acquires additional $\log \omega$ dressing and this additional dressing term contributes to the charge. This contribution (given in \eqref{A1u}) can be schematically written as $Q_{\text{1-loop}}\sim S_0\ h_{rr}^1$, where $S_0$ is the leading soft factor. 
\item
The two terms in the first line of \eqref{softthm2} constitute the classical soft factor. The low energy expansion of classical radiative field is controlled by the classical soft factors \cite{1808.03288, log waves}. The first term in the first line is related to the '$\hat{S}_1\ \overset{\text{class}}{h_{rr}^1}$'\footnote{$h_{rr}^1=\overset{\text{class}}{h_{rr}^1}+\overset{\text{quan}}{h_{rr}^1}$. Similarly $A_r^1=\overset{\text{class}}{A_r^1}+\overset{\text{quan}}{A_r^1}$.} term.  This part of the charge is directly related to the asymptotic acceleration of massless scalar particles under the gravitational force.  The second term in the first line is related to the '$S_0\ \overset{\text{class}}{h_{rr}^1}$' term. This part of the charge corresponds to the late time acceleration of the soft photon under gravitational force. The last two lines are absent in soft classical radiation and represent purely quantum effects.
\item
Let us switch off gravity for a moment and consider a purely electromagnetic setup. In \cite{1903.09133}, the authors discuss massive scalar particles in this setup. An interesting observation is that the classical part of the soft factor which is non zero for the massive case is absent for the massless case.  This result comes out naturally from the charge perspective also. The expected classical contribution is '$\hat{S}_1\ \overset{\text{class}}{A_r^1}$'. This classical mode i.e. $\overset{\text{class}}{A_r^1}$ given in \eqref{Arclassical} is trivial  and there is no classical contribution to charge (in absense of gravity). 
\item
It was noted in \cite{1808.03288} that if we assume the momenta of the hard particles is $\mathcal{O}(\hbar^0)$, neither the classical nor the quantum soft factor has any power of $\hbar$. Thus, an intriguing aspect of the 'quantum' terms is that these terms are independent of $\hbar$. These terms do not trivially vanish in classical limit ($\hbar\rightarrow 0$). In \cite{1903.09133}, the authors pointed out that there is a discontinuity in the quantized photon field in the limit $\omega\rightarrow 0$ and derived the 'quantum' part of $\log\omega$ coefficient from this discontinuity. Classical solutions are continous in $\omega\rightarrow 0$. In our case, discontinuities of the quantum photon and graviton fields contribute to $\overset{\text{quan}}{A_r^1}$ and $\overset{\text{quan}}{h_{rr}^1}$ respectively. This discontinuity is absent for massless scalar field. The last two lines of \eqref{softthm2} are obtained from $\hat{S}_1\ [\overset{\text{quan}}{h_{rr}^1}+\overset{\text{quan}}{A_r^1}]$ and $S_0 \ \overset{\text{quan}}{h_{rr}^1}$. 
\item For the massless case, the quantum contributions to the charge have divergent pieces arising from collinear configurations. Gravitational dressings of both massless scalar and photon fields have divergent pieces that cancel out in the total expression of the charge. The divergent part of electromagnetic dressing does not contribute to the charge. Thus, the charge is rendered finite.
\end{itemize}
%$A_r^1,\ h_{rr}^1$ admit contribution from a purely quantum modes in 

\section{Preliminaries}
We consider a theory with a massless scalar $\phi$ minimally coupled to  U(1) gauge field $A_\mu$  and gravitational field $g_{\mu\nu}$.  So, our system is described by the action : 
\begin{equation}
S = -\int d^{4}x \sqrt{-g}\ \Big[\frac{1}{4}g^{\mu\rho}g^{\nu\sigma}F_{\mu\nu}F_{\rho\sigma} + g^{\mu\nu}\left(D_{\mu}\phi\right)^{*}\left(D_{\nu}\phi\right)   +\ \frac{1}{2} R\Big],
\end{equation}
where $D_{\mu}\phi=\p_{\mu}\phi\ -\ ieA_{\mu}\phi$ and $8\pi G =1$. %where  and %$\nabla_{\mu}$ is the covariant derivative. 

We are interested in the asymptotic dynamics of above system. Massless particles end up at future null infinity ($r\rightarrow \infty$ with $t-r$ finite) which is represented as $\mathcal{I}^+$. To describe late time dynamics of massless fields, we need to use retarded co-ordinate system. The flat metric takes following form in this co-ordinate system ($u=t-r$) :
\be 
ds^2 = -du^2 - 2dudr + r^2\ 2\gamma_{z\bar{z}}\ dz d\bar{z}; \ \ \gamma_{z\bar{z}} = \frac{2}{(1+z\bar{z})^2}.\nn
\ee
We use $\hat{x}$ or $(z,\bar{z})$ interchangeably to describe points on $S^2$. An useful parametrisation of a 4 dimensional spacetime point is given by (Greek indices will be used to denote 4d Cartesian components) :
\be x^\mu = rq^\mu + u t^\mu,\ \ \  q^\mu=(1,\hat{x}), \ \ \  t^\mu=(1,\vec{0}).\label{q}\ee
Here, $q^\mu$ is a null vector that can be parameterised in terms of $(z,\bar{z})$ as $$q=\frac{1}{1+z\bar{z}}\lbrace1+z\bar{z},z+\bar{z},-i(z-\bar{z}),1-z\bar{z}\rbrace.$$ %$\mathcal{I}^+$ is described by the metric : 
%\be 
%ds^2 = -du^2  + r^2\gamma_{z\bar{z}}\ dz d\bar{z}.
%\ee
Dynamics of scalar is given by 
\begin{align}
&g^{\mu\nu}D_{\mu}D_{\nu}\ \phi(x)\  = 0. \label{eom}
\end{align}
Solution to this equation can be expanded around future null infinity.
Using stationary phase approximation, we can obtain the leading order coefficient in asymptotic expansion for massless scalars. It is given by \cite{qed1} 
\be \phi(u,r,\hat{x}) = \frac{1}{r}\ \phi^1(u,\hat{x})+...\ , \label{phi1}\ee
where \be \phi^1(u,\hat{x})= \frac{-i}{8\pi^2}\int d\omega\ [b(\omega,\hat{x})\ e^{-i\omega u} - d^\dagger(\omega,\hat{x})\ e^{i\omega u}\ ]. \label{phi}\ee
Next we turn to the gauge field. Choosing the covariant Lorenz gauge $\nabla_\mu A^\mu=0$, Maxwell's equations reduce to 
\be
\Box A_{\mu} =-j_\mu+j_{\mu}^{\text{grav}},\ \ \ j_{\mu} = ie\left(\phi D_{\mu}\phi^{*}-\phi^{\ast}D_{\mu}\phi\right). \ee $j_\sigma^{\text{grav}}$ represents the gravitational corrections and will be analysed in section \ref{phdress}. It should be noted that $A_\mu$ is used to denote the full solution including the homogenous part and inhomogenous terms coming from the U(1) current as well as gravitational coupling.

Using the fall offs given in \eqref{phi1} for massless scalars, we get following asymptotic behaviour for the current components :
\be  j_u = \frac{j_u^2(u,\hat{x})}{r^2} + ..., \ \ \ \ j_A = \frac{j_A^2(u,\hat{x})}{r^2} + ..., \ \ \ \ j_r = \frac{j_r^4(u,\hat{x})}{r^4} + ..., \ \ \ \ (A=z,\bar{z}).  \label{cur}\ee 
Above and henceforth, we denote the vector components on $S^2$ by capital latin alphabets. The asymptotic expansion of gauge field components that is consistent with above sources is given by :
\begin{align}
&A_{r}=\frac{A_{r}^1(\hat{x})}{r} + A_{r}^{\log}(u,\hat{x})\frac{\log r}{r^2} +...\ , &A_{u}=A_{u}^{\log}(u,\hat{x})\frac{\log r}{r} +\frac{A_{u}^1(u,\hat{x})}{r} + ...\ ,\ \ \ \ \nn\\
  &A_{A}= A^0_{A}(u,\hat{x}) +A_{A}^{\log}(u,\hat{x})\frac{\log r}{r} + ...\ .& \label{Afalloff}
\end{align}

Next let us consider the asymptotic behaviour of the gravitational field. We will work in the perturbative linear gravity regime where gravitational dynamics is confined to perturbations around flat space time : $g_{\mu\nu}=\eta_{\mu\nu}+h_{\mu\nu}$. In the de Donder gauge $\p_\mu \h^{\mu\nu}=0$, where $\bar{h}_{\mu\nu}=h_{\mu\nu}-\frac{1}{2}\eta_{\mu\nu}h^\sigma_{\ \sigma}$. The metric field satisfies $\Box \h_{\mu\nu} =-2T_{\mu\nu}$. The metric field admits following expansion : \footnote{Some of the coefficients are independent of $u$, this follows from the de Donder gauge condition itself.}
\begin{align}
&h_{rr}=\frac{h_{rr}^1(\hat{x})}{r}+h_{rr}^{\log}(u,\hat{x})\frac{\log r}{r^2} + ...\ , &h_{ur}=\frac{h_{ur}^1(u,\hat{x})}{r}+h_{ur}^{\log}(u,\hat{x})\frac{\log r}{r^2}  + ...\ ,\ \nn\\
 &h_{uu}=h_{uu}^{\log}(u,\hat{x})\frac{\log r}{r} +\frac{h_{uu}^1(u,\hat{x})}{r} + ...\ ,& h_{rA}=h_{rA}^0(\hat{x}) +h_{rA}^{\log}(u,\hat{x})\frac{\log r}{r^2}  + ...\ ,\nn\\
 &h_{uA}=h_{uA}^0(u,\hat{x})+h_{uA}^{\log}(u,\hat{x})\frac{\log r}{r}   + ...\ , \ \ \ \ \ &h_{AB}=r\ h^{-1}_{AB}(u,\hat{x}) +\log r\ h^{\log}_{AB}(u,\hat{x}) + ...\ . &\label{h}
\end{align}
There exists similar asymptotic expansion of fields at past null infinity ($r\rightarrow \infty$ with $t+r$ finite) represented by $\mathcal{I}^-$.
\subsection{Asymptotic conservation laws}
Classical equations of motion can be used to derive conservation laws of the form :
\be
Q^+[\lambda^+]\ |\ _{\mathcal{I}^+_-}\ \  = \ \ Q^-[\lambda^-]\ |\ _{\mathcal{I}^-_+}.
\ee
Here, $\mathcal{I}^+_-$ is the $u\rightarrow-\infty$ sphere of $\mathcal{I}^+$ and $\mathcal{I}^-_+$ is the $v\rightarrow\infty$ sphere of $\mathcal{I}^-$. The quantum versions of above statements are related to soft theorems. Here, $\lambda^+$ refers to an arbitrary parameter defined at $\mathcal{I}^+_-$. The parameter at $\mathcal{I}^-_+$ is related to it via antipodal map $\lambda^+(\hat{x})=\lambda^-(-\hat{x})$. Thus, we have a conservation law for every possible choice of $\lambda$. In this section, we will review the leading conservation law for QED. The asymptotic expansion given in \eqref{Afalloff} leads to following fall offs for the field strength : 
\begin{align}
& F_{ru}  =\frac{F^2_{ru}(u,\hat{x})}{r^2}+...\ , \ \ \ & F_{uA}  =F^0_{uA}(u,\hat{x})+...\ , &&\nn\\ & F_{AB}  =F^0_{AB}(u,\hat{x})+...\ ,
&F_{rA} \footnotemark =\frac{F^2_{rA}(u,\hat{x})}{r^2}+...\ . &&\label{Ffall}
\end{align}
\footnotetext{$F_{rA}$ actually starts at $F_{rA}=F^{\log}_{rA}(u,\hat{x})\frac{\log r}{r^2}+...$ due to presence of massless fields. Maxwell's equations imply $\p_u F_{rA}^{\log}=0$, so we set this mode to 0.}
The Maxwell's equations are given by $\nabla^{\nu}F_{\sigma\nu}\ =\ j_{\sigma}$ and imply following equations for the coefficients in \eqref{Ffall} :
\begin{align}
\partial_uF^2_{ru}\ +\  \partial_uD^BA_B^0\ =\ j_u^2,\nn\\
\partial_uF^2_{rA}\ -\ \frac{1}{2}\partial_AF^2_{ru}\ +\frac{1}{2} D^BF_{AB}^0\ =\ \frac{1}{2}j_A^2.\label{maxwell}
\end{align}
Let us use above equations to the study the $u$-behaviour of the field strength components $F^2_{ru}$ and $F_{rA}^2$. Around $|u|\rightarrow\infty$, the currents die stronger than any power law of $u$, so we can ignore the currents at $|u|\rightarrow\infty$. The large $u$ behaviour of $F_{ru}^2$ is thus fixed by $A_A^0$. Let us consider profiles of $A_A^0$ that are consistent with the tree level soft theorems. At tree level, the soft expansion is given by : $\tilde{A}_A^0 \sim \sum_{m=-1}^\infty S_m\ \omega ^m $. Using Fourier transform, we get following behaviour for radiative data :
\begin{align}
A_A^0|_{\mathcal{I}^+_-}\ =\ A_A^{0,0}(\hat{x}) u^0\  +\ ...\ . \label{urad}
\end{align}
'$...$' denote terms that fall off faster than any power law in $u$. Hence, around $u \rightarrow -\infty$, the field strength components admit following behaviour :
\begin{align} F^2_{ru} |_{\mathcal{I}^+_-}&=  u^0\ F^{2,0}_{ru}(\hat{x})\  + ... \ , \nn\\
F^2_{rA}|_{\mathcal{I}^+_-} &= u\ F^{2,-1}_{rA}(\hat{x})\ +\ u^0\ F^{2,0}_{rA}(\hat{x})\ + ...\ . \label{FrA}\end{align}

 In \cite{1703.05448}, for specific physical processes, the author showed that following relation holds between asymptotic values of the fields  :
\be
F^{2,0}_{ru}(\hat{x})\ | _{\mathcal{I}^+_-}\ \  = \ \ F^{2,0}_{rv}(-\hat{x})\ |_{\mathcal{I}^-_+}.
\ee
Above statement can be rewritten as conservation law for charges parameterized by a scalar function $\lambda$ :
\be
Q_{\text{lead}}^+[\lambda^+]\ |_{\mathcal{I}^+_-}\ \  = \ \ Q_{\text{lead}}^-[\lambda^-]\ | _{\mathcal{I}^-_+}.\label{c1}
\ee 
$Q_{\text{lead}}^+[\lambda^+]=\int d^2z\ \lambda^+(\hat{x})\ {F^{2,0}_{ru}}(\hat{x})$. $Q_{\text{lead}}^-$ is defined analogously. And $\lambda^+(\hat{x}) = \lambda^-(-\hat{x})$.  This conservation law was proved for generic processes in \cite{1703.07884}. In \cite{1703.07884}, the authors studied the Maxwell's equations at spatial infinity and showed that \eqref{c1} holds generically for any solutions of Maxwell's equations. The charge in \eqref{c1} is in fact the charge associated to the large U(1) gauge transformations. The Ward identity for this charge is equivalent to the leading soft photon theorem\cite{qed1,1703.05448}.

\subsection{Outline of the paper}
In this paper, we construct the asymptotic charge corresponding to $\log \omega$ soft theorem for massless scalar QED coupled to gravity. To study the $\log \omega$ soft theorem we need to incorporate the effect of long range forces on asymptotic dynamics. In absence of long range forces, asymptotic fields satisfy free equations of motion. Including the correction to the asymptotic dynamics due to long range interactions leads to dressing of the free fields. 
\begin{itemize}
\item In sections 3 and 4, we discuss dressing of massless scalar field and photon field respectively. We show that as a result of these dressings, the $u$-fall offs given in \eqref{urad} are corrected to 
\be
A_A^0|_{\mathcal{I}^+_-}\ =\  A_A^{0,0}(\hat{x})  u^0\ + A_A^{0,1}(\hat{x})  \frac{1}{u}\  + \ \mathcal{O}(\frac{1}{u^2})\ .\label{1/u}
\ee
The $\frac{1}{u}$ term which is absent in \eqref{urad} is a direct consequence of long range interactions between particles. This late time mode implies that there is a new mode in the soft expansion of the gauge field namely the $\log \omega$ mode. The exact contribution from scalar dressing to $1/u$ term is given in \eqref{Ascalar}. This term is due to long range forces acting on the massless scalar particles. Similarly, the soft photon itself would feel the effect of long range gravitational force. This contribution from gravitational dressing of photon to $1/u$ term is given in \eqref{photon dressing}. 
\end{itemize}
Using \eqref{maxwell}, it can be shown that the $1/u$ leads to a $\log u$ term in \eqref{FrA}. The precise relation between these modes is given in \eqref{Ffut}. Thus, the $u$-behaviour of $F_{rA}^2$ is given by : 
\be
F^2_{rA}|_{u \rightarrow -\infty} = \ u\ F^{2,-1}_{rA}(\hat{x})\ +\ \log (-u)\ F^{2,\log}_{rA}(\hat{x})\ + ... \ . \label{1}\ee
A similar analysis at the past shows that there are new modes in the field due to the long range forces acting on the incoming particles. We show in \eqref{pastF} of Appendix A that expansion around the past null infinity is modified to : 
\be
F_{rA}|_{v \rightarrow \infty} =  \frac{\log r}{r^2}\ [v^0 \ F^{\log,0}_{rA}(\hat{x})\ +...]+ \mathcal{O}(\frac{1}{r^2}).\label{2}\ee
Here, '$...$' denote terms that fall off faster than power law in $v$. This $\log r$ mode was missed in \cite{1903.09133}. Hence, the conservation law proposed by \cite{1903.09133} is not entirely correct\footnote{We thank Arnab Priya Saha and Biswajit Sahoo for discussions about this point.}. Let us propose following conservation law for the logarithmic modes\footnote{We thank the authors of \cite{1903.09133} for suggesting this new conservation law.}: 
\be
{F^{2,\log}_{rA}}(\hat{x})\ |_{\mathcal{I}^+_-}\ \  = \ \   {F^{\log,0}_{rA}}(-\hat{x})\ |_{\mathcal{I}^-_+}. \label{cons}
\ee
 We have checked \eqref{cons} explicitly for classical processes with no incoming radiation. We believe that above conservation law can be proved by following the analysis of \cite{1703.07884,1810.04619}. In this paper we will assume this law and construct the associated charges and then prove that these charges reproduce the $\log\omega$ soft theorem.
%At tree level, in presence of massive fields the field strength components at future null infinity admit fall offs similar to \eqref{FrA} i.e.  
%\be
%F_{rA}|_{u \rightarrow -\infty} = \frac{1}{r^2}\ [\ u\ F^{2,-1}_{rA}(\hat{x})\ +\ u^0\ F^{2,0}_{rA}(\hat{x})\ + u^{-\infty}\ ] + ...\ , \ee
\begin{itemize}
\item In section 5, we start with above conservation law and identify the soft and hard modes of the charge. We refer to it as 1-loop charge, since it is expected to be related to the $\log \omega$ soft theorem which is 1-loop exact. The expression of soft charge is given in \eqref{soft1}. This operator isolates soft $\log \omega$ mode of the photon field. The expression of hard charge is given in \eqref{hard} and \eqref{1/ufull}. As discussed in Section 1, the hard charge is given in terms of dressings $h_{rr}^1$ and $A_r^1$.  
\item To evaluate the action of the hard charge, we find the expression of $h_{rr}^1$ and $A_r^1$ modes in section 6. Each of these modes has a classical and a quantum part : $h_{rr}^1=\overset{\text{class}}{h_{rr}^1}+\overset{\text{quan}}{h_{rr}^1}$ and $A_r^1=\overset{\text{class}}{A_r^1}+\overset{\text{quan}}{A_r^1}$. In section 6.1, the classical modes are obtained by evolving the sources with retarded propagator. $\overset{\text{class}}{A_r^1}$ given in \eqref{Arclassical} turns out to be trivial. This implies that classical part of electromagnetic contribution to the charge vanishes. \\
The quantum modes are slightly subtle. These modes are directly related to the discontinuity of fields in $\omega\rightarrow 0$.  As a result of this discontinuity in the graviton field, we get a $\log u$ term in $C_{AB}$ as seen in \eqref{Czz log} and \eqref{Cbarz log}. These $\log u$ terms contribute to $\overset{\text{quan}}{h_{rr}^1}$ via \eqref{hrr quan1}. $\overset{\text{quan}}{A_r^1}$ is obtained similarly.
\item Having constructed the full charge, in section 7, we finally write down the Ward identity for the 1-loop charge and show that it is equivalent to the Sen-Sahoo soft theorem.
\end{itemize}

\section{Dressing of  massless scalar field}
In this section we study the dressing of free scalar fields under the effect of long range forces and find the resultant correction to the asymptotic field. In particular, we will show that long range forces produce a new mode in \eqref{urad} that falls off as $1/u$ . 

%In absence of long range forces, particles are free asymptotically. 
For massive fields the effect of long range forces is obtained perturbatively by studying asymptotic potential order by order around $t\rightarrow \infty$. This leads to the well known Faddeev-Kulish dressing of massive scalars\cite{FK}. For massless scalars, the asymptotic states live at null infinity. So, we will study the corrections to the free equation of motion at null infinity. 
Massless scalars satisfy following equation :
\begin{align}
&g^{\mu\nu}D_{\mu}D_{\nu}\ \phi(x)\  = 0. \label{esca}
\end{align}
Let us expand above equation around future null infinity. 
Using the fall offs given in \eqref{Afalloff} and \eqref{h}, we find that the leading order equation is (at $\mathcal{O}(\frac{1}{r^2})$) :
\be -2\p_u\p_r\phi -\frac{2}{r}\p_u\phi= \frac{h_{rr}^1(\hat{x})}{r}\p_u^2\phi-2ie\frac{A_{r}^1(\hat{x})}{r}\p_u\phi. \label{ephi}\ee
Thus, the leading order effect of long range forces on the massless field is given by $h_{rr}^1$ and $A_r^1$. The solution of above equation is given by :
\be \phi(x)= -\frac{ie^{ieA^1_r(\hat{x})\log\frac{r}{r_0} }}{8\pi^2r}\int d\omega\ [b(\omega,\hat{x})\ e^{-i\omega u}e^{i\omega\log \frac{r}{r_0} \frac{h^1_{rr}(\hat{x})}{2}\ } - d^\dagger(\omega,\hat{x})\ e^{i\omega u}e^{-i\omega\log \frac{r}{r_0} \frac{h^1_{rr}(\hat{x})}{2}\ }], \label{phidress}\ee
where, $b$ and $d^\dagger$ are the free data for massless scalar. $r_0$ depends on scales of short range interactions, hence $r_0<<r$. For our analysis we can set $r_0=1$ to avoid clutter. On quantisation, $b$ can be interpreted as the annihilation operator for free particles while $d$ would become the annihilation operator for free antiparticles (see \eqref{phi}). From \eqref{phidress}, we see that the leading order effect of long range forces is to associate a cloud of photons and gravitons to a free massless scalar particle. These dressing factors ($h_{rr}^1$ and $A_r^1$) are analogous to the  Fadeev-Kulish dressing of a free massive scalar particle.  Next we find the correction to the U(1) current as a result of long range forces. Dressing of scalar field leads to a new logarithmic fall off in the current \eqref{cur} : 
\be j_A\ =\ \ {j^{\log}_A}\ \frac{\log r}{r^2}+ \frac{{j}^2_A}{r^2} \ + ...\ ,\ee
where
\begin{align}
 {j^{\log}_A} 
&=  -\ \frac{1}{2}\p_A h_{rr}^1\  j_u^2 +2e^2\  \p_A A_r^1\ |\phi^1 |^2.\label{logjA}
\end{align}
Let us find the the gauge field generated by these new logarithmic fall offs in the current. In Lorenz gauge, we have $\Box A_\mu = -j_\mu$. (This equation admits corrections due to gravity; gravitational corrections will be analysed in section \ref{phdress}). Using the retarded propagator, the solution to the gauge field is given by :
\begin{align}
A_\sigma(x)= \frac{1}{2\pi }\int d^4x'\ \delta(\ (x-x')^2)\  \Theta(t-t')\ j_\sigma(x').
\end{align}
We will substitute the new logarithmic modes of the current in above expression and find the resultant contribution to the field. The details of the calculation have been relegated to Appendix \ref{AA}. We show that the log modes give rise to a $1/u$ term in $A_A^0$ such that the coefficient is given by \eqref{Ascalar1} :
\begin{align}
A^{0,1}_{\bar{z}}(\hat{x})|_\text{scal}
&= \frac{1}{4\pi }\frac{\sqrt{2}}{1+z\bar{z}}\int_{-\infty}^\infty du'\int _{S^2} d^2z'\ \frac{\epsilon_-^\mu q^\sigma}{q.q'}\ q'_{[\mu}D'^A q'_{\sigma]}\  {j^{\log}_A}.\label{Ascalar}
\end{align}
We have added a subscript 'scal' to highlight the fact that this contribution arises from scalar field dressing. This $1/u$-term has been discussed in the context of scattering of point particles in \cite{log memory em}. In above expression we have used the following basis for polarisation vectors \cite{qed1}:
\be \epsilon^\mu_- = \frac{1}{\sqrt{2}}\frac{\p}{\p \bar{z}} [(1+z\bar{z})q^\mu],\ \  \epsilon^\mu_+ = \frac{1}{\sqrt{2}}\frac{\p}{\p z} [(1+z\bar{z})q^\mu].\label{pol}\ee
The expression for $A_z$ can be obtained from the expression for $A_{\bar{z}}$ by replacing $\epsilon_-$ by $\epsilon_+$. 

Let us recall \eqref{urad}, this expression describes the $u$-fall offs of $A_A^0$ when the long range forces were ignored. The $u^0$ term in this expression is related to the leading soft theorem. This term is unchanged by long range forces which reflects the fact that the leading soft theorem does not receive any loop corrections. We see that including the effect of long range forces introduces a new $1/u$-term given in \eqref{Ascalar} that is absent in \eqref{urad}. The $1/u$ term is $\mathcal{O}(e^3)$ (or $\mathcal{O}(eG)$ for gravitational correction). Thus the soft expansion of the gauge field changes non-trivially as we go to higher order in couplings. 
 %This fact will play a role in definition of hard charge.\\
%So, to summarise we have studied the $\log r$ dressing of massless scalars and showed that it leads to a $1/u$ term in $A_{A}$.

%To summarise, in this section we described how the dressing of the massless scalar field contributes to logarithmic modes in the asymptotic gauge field. In \eqref{cons}, we have proposed that these logarithmic modes obey an asymptotic conservation law. Eventually we will study the Ward identity corresponding to this charge. Prior to that we study the effect of the long range forces on the gauge field in Section 4. It turns out that the logarithmic modes in \eqref{1} and \eqref{2} receive contribution from dressing of the gauge field as well.

%We will study this conservation law in presence of massless scalars and dynamical gravity. So, we will first identify logarithmic fall-offs that arise due to dressing of massless scalars and photons\footnote{ Dressing of gravitons does not affect the Ward identity for photons.} and contribute to the conservation equation. The loop level soft theorem can then be derived from this conservation equation.

%In the next section, we will study dressing of massless scalar and identify the resultant $1/u$ term in $A_{A}^0$ that eventually contributes to \eqref{cons}.

\section{Dressing of  gauge field} \label{phdress}
 In this section, we study the effect of long range gravitational force on gauge fields. Before delving into the calculation let us state our result. The leading order correction to \eqref{urad} as a result of coupling of photon with gravity is :  
\begin{align}
A^{0,1}_{\bar{z}}(\hat{x})|_\text{grav dress}
&= -\frac{1}{8\pi }\ \frac{\sqrt{2}}{1+z\bar{z}} \  h^1_{rr}(\hat{x})\ \lim_{\omega\rightarrow 0}\omega a_-(\omega,\hat{x}).\label{photon dressing}
\end{align}
For $A^{0,1}_{z}$, we have to replace negative helicity operators with positive helicity operators.
%This term will contribute to the $\frac{\log u}{r^2}$ mode at the future by virtue of \eqref{Ffut}.

Let us derive above expression. First we start with the homogenous equation $\Box A^{hom}_\mu =0$. Asymptotically such a solution exhibits following form \cite{qed1}:
\be A^{hom}_\sigma(u,r,\hat{x})= -\frac{i}{8\pi^2r}\int d\omega\ [a_\sigma(\omega,\hat{x})\ e^{-i\omega u} -a_\sigma^\dagger(\omega,\hat{x})\ e^{i\omega u}\ ],\label{hom}\ee
where $a_\sigma=\sum_{r=+,-}\epsilon^r_\sigma\ a_r$. Let us turn on the sources. Choosing the generalised Lorenz gauge $\nabla_\mu A^\mu=0$, Maxwell's equations reduce to : 
\be
\nabla^2 A_{\mu} =-j_\mu+{R_\mu}^\nu A_\nu.
\ee
$R_{\mu\nu}$ is the Ricci tensor. Let us ignore the U(1) current here. Since we are working in perturbative gravity, we will retain the leading order gravitational corrections. We get : 
\begin{align}
 \Box A_\sigma= j^{\text{grav}}_\sigma,
\label{boxA} 
\end{align}
where we have defined :
$$j^{\text{grav}}_\sigma=h^{\mu\nu}\partial_\mu\partial_\nu A_\sigma + \eta^{\mu\nu}\Gamma^\rho_{\mu\nu}\partial_\rho A_\sigma +2 \eta^{\mu\nu}\Gamma^\rho_{\mu\sigma}\partial_\nu A_\rho\ + \eta^{\mu\nu}\ A_\lambda\ \partial_\mu\Gamma_{\nu\sigma}^\lambda   + [\p_\mu \Gamma^\mu_{\nu\sigma}-\p_\nu \Gamma^\mu_{\mu\sigma}] A^\nu +\ \mathcal{O}(G^2).$$
Next $ j^{\text{grav}}_\sigma$ can be evaluated on the zeroth order solution. Using \eqref{h} and \eqref{hom}, we see that the source has following behaviour around future null infinity :
\be  j^{\text{grav}}_\sigma(x) = \frac{1}{r^2}h^1_{rr}\partial_{u}^2 A^1_\sigma\ +\  \mathcal{O}( \frac{1}{r^3}). \label{200}\ee
The $\mathcal{O}( \frac{1}{r^3})$ terms in $  j^{\text{grav}}_\sigma(x)$  produce subleading corrections, hence are not relevant for our analysis. 
Analogous to the massless scalar equation \eqref{ephi}, we get :
\be -2\p_u\p_rA_\sigma -\frac{2}{r}\p_uA_\sigma= \frac{h_{rr}^1(\hat{x})}{r}\p_u^2A_\sigma. \ee
The solution to above equation is given by :
\be A_\sigma(u,\hat{x})= -\frac{i}{8\pi^2}\int d\omega\ [a_\sigma(\omega,\hat{x})\ e^{-i\omega u}e^{i\omega\log r \frac{h^1_{rr}(\hat{x})}{2}\ } -a_\sigma^\dagger(\omega,\hat{x})\ e^{i\omega u}e^{-i\omega\log r \frac{h^1_{rr}(\hat{x})}{2}\ }]. \label{logr}\ee
Thus, the $\log r$ dressing of photons is exactly similar to the $\log r$ dressing of massless scalars. This dressing does not contribute to the loop level charge. The contribution to the loop level charge comes from $1/u$ term. So, we need to check if the source \eqref{200} induces a $1/ru$ in $A_{\mu}$.  Using the Green's function for D'Alembertian operator :
\begin{align}
A^{\text{grav}}_\sigma(x)=- \frac{1}{2\pi }\int d^4x'\ \delta_+(\ (x-x')^2)\ h^1_{rr}(z')\partial_{u'}^2 A^1_\sigma(u',z').\nn
\end{align}
We have used a superscript 'grav' to highlight the fact that this mode arises due to gravitational coupling. Taking the limit $r\rightarrow \infty$ with $u<r$ : 
\begin{align}
A^{\text{grav}}_\sigma(u,r,\hat{x})&=- \frac{1}{4\pi r}\int du' dr' d^2z'\ \delta_+(u'+r'-u-\vec{x'}.\hat{x})\  h^1_{rr}(z')\partial_{u'}^2 A^1_\sigma(u',z')\ +\ \mathcal{O}(\frac{1}{r^2}),\nn\\
&=- \frac{1}{4\pi r}\partial_u\Bigg[\int_{-\infty}^\infty du'\int_0^\infty dr'\int_{\mathcal{S}^2} d^2z'\ \delta(u'+r'-u-\vec{x'}.\hat{x})\  h^1_{rr}(z')\partial_{u'} A^1_\sigma(u',z')\ \Bigg].\nn
\end{align}
$\partial_{u'} A^0_\sigma$ vanishes for $|u'|>u_0$ as to the zeroth order the particles are free for $|u'|>u_0$, where, $u_0$ is some time scale that is set by short range interactions. We can use rotational symmetry to align $\hat{x}$ along $z'$-axis :
\begin{align}
A^{\text{grav}}_\sigma(u,r,\hat{x})&=- \frac{1}{2 r}\partial_u\Bigg[\int_{-u_0}^{u_0} du' \int_0^\infty dr' \int_{-1}^1d\cos\theta' \ \frac{1}{r'} \delta\big(\cos\theta'-1+\frac{u-u'}{r'}\big)\  h^1_{rr}(\theta')\ \partial_{u'} A^1_\sigma(u',\theta')\ \Bigg].\nn
\end{align}
We will use the delta function to do the $\theta'$ integral. $\cos\theta'\ \epsilon\  [-1,1] $ leads to a bound on other integration variables. There are two allowed ranges : $ u>u',\ 2r' > u-u' ; \ \ \ \ \ u'>u, \ 2r'<-(u'-u)$. The second range is inadmissible as $r'$ needs to be positive. Also, $r'$-integral needs to be regulated with some IR cutoff.
\begin{align}
A^{\text{grav}}_\sigma(u,r,\hat{x})
&= - \frac{1}{2 r}\ \partial_u\Big[\int_{-u_0}^{u_0} du' \int_ {\frac{u-u'}{2}}^R \frac{dr'}{r'} h^1_{rr}(\theta')\partial_{u'} A^1_\sigma(u',\theta')\ |_{\cos\theta'=1-\frac{u-u'}{r'}}  \Big].\nn
\end{align}
Taylor expanding the integrand around $\cos\theta'=1$, we get the leading order contribution in $u\rightarrow\infty$ limit to be :
\begin{align}
A^{\text{grav}}_\sigma(u,r,\hat{x})
&=  \frac{1}{ 2r}\ \Big[\ \int_{-u_0}^{u_0} du' \frac{1}{u-u'} h^1_{rr}(\theta')\partial_{u'} A^1_\sigma(u',\theta')\ |_{\cos\theta'=1}  \Big]. \label{25}
\end{align}
Above expression can be readily related to insertion of leading soft mode :
\begin{align}
A^{\text{grav}}_\sigma(u,r,\hat{x})_{u\rightarrow \infty}
&= \frac{1}{ 2r}\frac{1}{u}\  h^1_{rr}(\hat{x})\ \int_{-u_0}^{u_0} du' \partial_{u'} A^1_\sigma(u',\hat{x}),\nn\\
&= -\frac{1}{8\pi r}\frac{1}{u}\  h^1_{rr}(\hat{x})\ \lim_{\omega\rightarrow 0^+}\omega \ a_\sigma(\omega,\hat{x}).\label{ph dressing}
\end{align}
One point needs to be highlighted here. We have defined the soft limit such that $\omega$ is taken to 0 from the positive side. This definition is consistent with the fact that we have used retarded propagator in our derivation. We have also derived the results of this paper using Feynman propagator (whence we need to use a symmetric notion of soft limit). This alternative derivation involving Feynman propagator is not included in this paper, we will discuss it elsewhere.

Above expression can be combined with \eqref{logr} to arrive at  : 
\be A_\sigma(u,r,\hat{x})= -\frac{i}{8\pi^2r}\int d\omega\ [a_\sigma(\omega,\hat{x})\ e^{-i\omega u}e^{i\omega\log (r\omega) \frac{h^1_{rr}(\hat{x})}{2}\ } -a_\sigma^\dagger(\omega,\hat{x})\ e^{i\omega u}e^{-i\omega\log (r\omega) \frac{h^1_{rr}(\hat{x})}{2}\ }]. \label{Adress}\ee
\eqref{Adress} is the main result of this section. This expression represents the effect of gravitational field on outgoing photons. In presence of incoming soft photon, above analysis needs to be repeated at past null infinity. The precise contribution to the loop charge is given in \eqref{photon dressing}. We can arrive at \eqref{photon dressing} by co-ordinate transformation of \eqref{ph dressing}. We also note that similar to massless scalars, the gravitational contribution to the charge depends only on $h_{rr}^1$.  
%Classical part of these expressions is simpler and we discuss it in Appendix  \ref{A}. In the next section, Next we need expressions for $h_{rr}^1$ and $A_r^1$.

\section{ The Asymptotic charge}
%Finally we have all the ingredients to write down the loop level Ward identity. 
In this section we will obtain the explicit expression for the 1-loop asymptotic charge. We start with the conservation equation \eqref{cons} : 
\be
{F^{2,\log}_{rA}}(\hat{x})\ \  = \ \   {F^{\log,0}_{rA}}(-\hat{x}).
\ee
We recall that the LHS is the coefficient of the $\frac{\log u}{r^2}$-mode present at the future. Similarly the RHS is the coeffficient of the $\frac{\log r}{r^2}$ mode living at the past. We multiply above equation with an arbitrary parameter $V^A$ and integrate over the sphere to get
\be
\int_{\mathcal{I}^+_-}d^2z\ V^A(\hat{x}){F^{2,\log}_{rA}}(\hat{x})\ \  = \ \  \int_{\mathcal{I}^-_+}d^2z\ V^A(-\hat{x}) {F^{\log,0}_{rA}}(-\hat{x}).\label{cons1}
\ee
The charge at the future is defined by $Q_{\text{1-loop}}^+[V^A_+]=-\int  d^2z\ V_+^A F^{\log,0}_{rA}\ |\ _{\mathcal{I}^+_-}$. The past charge is defined similarly. Our claim is that this conservation law reproduces the outgoing soft photon theorem given in \eqref{softthm2}. In the most general scenario there exists a $\log v$ mode at past. The $\log v$ mode corresponds to incoming soft photon and we have set these modes to zero. Similar conservation law that relates $\log v$ mode at $ {\mathcal{I}^-_+}$ to $\log r$ mode at $ {\mathcal{I}^+_-}$ reproduces the incoming soft theorem.  

Let us study the future charge : 
\begin{align}
Q^{\text{1-loop}}_+[V] &=- \int  d^2z\ V^A {F^{2,\log}_{rA}}\ |\ _{\mathcal{I}^+_-},\   \nn\\
&=u^2 \partial_u^2 \int  d^2z\ V^A {F^2_{rA}}\ |\ _{u\rightarrow -\infty}.\   \nn
\end{align}
The $u$-operator isolates the coefficient of the log $u$ term of $F^2_{rA}$. We can rewrite the future charge as an integral over entire future null infinity minus the term at $\mathcal{I}^+_+$ . 
\begin{align}
Q_+^{\text{1-loop}}[V]
&=\  -\int_{-\infty}^\infty du'\int d^2z\  V^A  \partial_u\ [u^2 \partial_u^2 {F}^2_{rA}] -\ \int  d^2z\ V^A F^{2,\log}_{rA}\ |\ _{\mathcal{I}^+_+},\nn\\
 &: =Q_+^{\text{soft}}[V]+Q_+^{\text{hard}}[V].\label{Q}
\end{align}
This defines the soft and hard parts of asymptotic charge. 
We can simplify the soft charge expression further. Using Maxwell's equation \eqref{A comp} for $\partial_u {F}^2_{rA}$, we get :
\begin{align}
Q_+^{\text{soft}}&=-\frac{1}{2}\int_{-\infty}^\infty du'\int  d^2z'\  V^A  \partial_u\ \big[u^2 \partial_u[\partial_A{F}^2_{ru} - D^BF_{AB}^0+j_A^2]\ \big]+...\ .\label{Qs}
\end{align}
In above expression '...' arise due to the gravity corrections to Maxwell's equations. We have studied these terms explicitly in Appendix \ref{A} and we show that these corrections vanish. $j_A^2$ does not have a $1/u$-term, so $j_A^2$ also drops out of above expression and we get :
\begin{align}
Q_+^{\text{soft}}
&=\int_{-\infty}^\infty du'\int  d^2z'\ \Big[ V^z(\hat{x}')  \partial_u\ [u^2 \partial_u D_z D^{\bar{z}}A^0_{\bar{z}}(u,\hat{x}') ]+ z' \leftrightarrow \bar{z}'\Big],\nn\\
&=\int_{-\infty}^\infty du'\int d^2z'\ \Big[ D_{z}'^2 V^z \ \gamma^{z\bar{z}}\  \partial_u\ [u^2 \partial_u A^0_{\bar{z}}(u,\hat{x}') ]+ z' \leftrightarrow \bar{z}'\Big].
\end{align}
The last line was derived using integration by parts. Next it is instructive to go to the frequency space :
\begin{align}
Q_+^{\text{soft}}&=\int   d^2z'\ \big[  D_{z}'^2 V^z \ \gamma^{z\bar{z}}\ \lim_{\omega\rightarrow 0^+}  \omega\ \partial_\omega^2\ \omega\ \tilde{A}_{\bar{z}}^0(\omega,\hat{x}') + z' \leftrightarrow \bar{z}'\Big]. \nn
\end{align}
As we discussed at the end of section 4, we have defined the soft limit from positive side. The gauge field can be expressed in terms of Fock operators as : 
\be \tilde{A}_{\bar{z}}^0(\omega,\hat{x}) = -i\sqrt{2}\frac{a_{-}(\omega,\hat{x})}{4\pi(1+z\bar{z})} \ \ ...\ \ \omega >0,\ \ \ \tilde{A}_{\bar{z}}^0(\omega,\hat{x}) = i\sqrt{2}\frac{a^\dagger_{+}(-\omega,\hat{x})}{4\pi(1+z\bar{z})} \ \ ...\ \ \omega <0.\ee
So we get :
\begin{align}
Q_+^{\text{soft}}
&=-\frac{i}{4\pi}\int   d^2z'\  \Big[D_{z}'^2 V^z \ \sqrt{\gamma'^{z\bar{z}}} \lim_{\omega\rightarrow 0^+}  \omega\ \partial_\omega^2\ \omega\ \ a_{-}(\omega,\hat{x}')  + z' \leftrightarrow \bar{z}'\Big].\label{soft1}
\end{align}
Thus, the action of $Q_+^{\text{soft}}$ involves insertion of zero energy photon modes. The $\omega$-derivatives in particular isolate the coefficient  of soft $\log \omega$ mode. 

Next let us turn to the expression of future hard charge :
\begin{align}
Q_+^{\text{hard}}&=-\int d^2z'\  V^A \ F^{2,\log}_{rA}(\hat{x}').\nn
\end{align}
Using \eqref{maxwell}, we have \be \partial^2_uF^2_{rA}\ +\ \frac{1}{2}\partial_u\p_AD^BA_B^0\ +\frac{1}{2} \p_uD^BF_{AB}^0\ =\ \frac{1}{2}\p_uj_A^2.\ee
From above equation we get the precise relations :
\be
{F^{2,\log}_{rz}} =-\gamma^{z\bar{z}}\ D_z^2 {A^{0,1}_{\bar{z}}}\text{  and  } {F^{2,\log}_{r\bar{z}}} =-\gamma^{z\bar{z}}\ D_{\bar{z}}^2 {A^{0,1}_{z}}. \label{Ffut}
\ee
We recall that $A_A^{0,1}(\hat{x})$ denotes following mode in the gauge field : $A_A(x)\sim A_A^{0,0}(\hat{x})+A_A^{0,1}(\hat{x}) \frac{1}{u}+...\ $ . Using \eqref{Ffut} in the expression for the hard charge, it can be written in terms of coefficient of the $\frac{1}{u}$ mode. 
\begin{align}
Q_+^{\text{hard}}&=\int d^2z'\  V^{z}\ \gamma^{z\bar{z}}\ D^2_z {A^{0,1}_{\bar{z}}}(\hat{x}') +\int d^2z'\  V^{\bar{z}} \gamma^{z\bar{z}} \ D^2_{\bar{z}} {A^{0,1}_{z}}(\hat{x}') 
%&=\int d^2z'\ D_z'^2 V^z(z,z') \ \overset{0,1}{A_{\bar{z}}}(z')+\int d^2z'\ D_{\bar{z}}'^2 V^{\bar{z}}(z,z') \ \overset{0,1}{A_{z}}(z').\ 
\end{align}
 To avoid unnecessary cluttering of equations we will work with $V^{\bar{z}} = 0$. Then we can integrate by parts to get following equation :
\begin{align}
Q_+^{\text{hard}}&=\int d^2z'\ D_{z}'^2 V^z \ \gamma^{z\bar{z}}\ {A^{0,1}_{\bar{z}}}(\hat{x}'),\label{hard}
\end{align}
Using \eqref{Ascalar}, \eqref{logjA} and \eqref{photon dressing} we get :
\begin{align}
A^{0,1}_{\bar{z}}(\hat{x})&=\ \frac{\sqrt{\gamma_{z\bar{z}}}}{4\pi } \int_{-\infty}^\infty du'\int { d^2z'}\  \frac{q^\mu\  \epsilon^\sigma _-}{q.q'}\ q'_{[\sigma}\p'_{q^\mu]}\ [- \ \frac{1}{2}h_{rr}^1(\hat{x}')j_u^2(\hat{x}')\ + 2 e^2A_r^1(\hat{x}')|\phi^1(\hat{x}')^2\ ]\nn\\
&-\frac{\sqrt{\gamma_{z\bar{z}}}}{8\pi } \ h^1_{rr}(\hat{x})\ \lim_{\omega\rightarrow 0}\omega a_-(\omega,\hat{x}).\label{1/ufull}
\end{align}
Equations \eqref{hard} and \eqref{1/ufull} provide us the expression of the future hard charge. We recall that the first line is a result of long range forces acting on the massless scalar field while the second line is due to the gravitational force acting on soft photons.

Let us turn to the the expression of the past charge. We have :
\begin{align}
Q^{\text{1-loop}}_-[V] &=- \int  d^2z\ V^A F^{\log,0}_{rA}\ |\ _{\mathcal{I}^-_+}.\   \nn
\end{align}
We know from \eqref{pastF} that $F_{rA}^{\log,0}$ depends only on particle currents i.e. it has no contribution from radiation. Thus, at past the charge is entirely  made of hard modes.
\begin{align}
Q^{\text{1-loop}}_-[V] &=- \int  d^2z\ V^A {F^{\log,0}_{rA}}\ |\ _{\mathcal{I}^-_+}\  \ \ :=\ Q_-^{\text{hard}}[V].  \nn
\end{align}
Thus as we had mentioned earlier we see that the conservation law that we have started with in \eqref{cons1}, reproduces outgoing soft theorem. An analogous conservation law that relates $\log v$ mode at $ {\mathcal{I}^-_+}$ to $\log r$ (a purely hard mode) at  $ {\mathcal{I}^+_-}$ will reproduce the incoming soft theorem.

Using \eqref{pastF}, the charge at past can be recast as :
\begin{align}
Q_-^{\text{hard}}&=-\int d^2z'\ D_{z}'^2 V^z \ \gamma^{z\bar{z}}\ {B^{\log}}(\hat{x}'),\label{hardp}
\end{align}
where,
\be
{B^{\log}}(\hat{x}) = \frac{1}{4\pi}\frac{\sqrt{2}}{1+z\bar{z}} \int_{-\infty}^\infty dv'\int_{S^2} d^2z'\ \frac{q^\sigma\epsilon_-^\mu }{q.q'}\ q'_{[\mu}\p'_{q^\sigma]}\ [- \ \frac{1}{2}h_{rr}^1(-\hat{x}')j_u^2(-\hat{x}')\ + 2 e^2A_r^1(-\hat{x}')|\phi^1(-\hat{x}')^2\ ].\label{Blog}
\ee
 %It is to be noted that $ {B}^{\log}$ is exactly the same operator as the part of ${A^{0,1}_{\bar{z}}}$ originating from scalar field dressing that is given in \eqref{Ascalar}. 

\section{Expressions for $h_{rr}^1$ and $A_r^1$}
In the preceding section, we studied the expression of the 1-loop asymptotic charge. The hard charges depend on $h_{rr}^1$ and $A_r^1$ via \eqref{1/ufull} and \eqref{Blog}. In this section we will compute the expressions for $h_{rr}^1$ and $A_r^1$.
\subsection{Classical part}
We know that the solution for gauge field in Lorenz gauge is given by :
\begin{align}
A_{\mu}(x^\mu)|_{class}&= \frac{1}{2\pi}\int d^4 x'\ \delta\big(\ (x-x')^2\ \big)\Theta(t-t')\ j_{\mu}(x'),\nn
\end{align}
where we have used the retarded propagator. We need the $\frac{1}{r}$ component of above expression to find classical part of $A_r^1$. Taking large $r$ limit we get :
\begin{align}
A_{\mu}(u,r,\hat{x})|_{class}&= -\frac{1}{4\pi r}\int_{-\infty}^\infty du'\int d^2z'\ \frac{j^2_{\mu}(\hat{x}',u')}{q.q'}.\nn
\end{align}
It can be checked that above expression is consistent with the fall offs mentioned in \eqref{Afalloff}. In particular we have :
\begin{align}
\overset{\text{class}}{A_{r}^1}(\hat{x})&= \frac{1}{4\pi r}\int_{-\infty}^\infty du'\int d^2z'\  j^2_{u}(\hat{x}',u').\label{Arclassical}
\end{align}
This part of $A_{r}^1(x)$ is just a constant (i.e. independent of $u,\hat{x}$) hence does not contribute to the hard charge as seen from \eqref{1/ufull}. Thus, the classical electromagnetic dressing is trivial. This is consistent with the absence of classical $\log \omega$ term in soft electromagnetic radiation (in absence of gravitational coupling)\cite{1808.03288}.

Similarly, in De-Donder gauge, the metric perturbations satisfy $\Box \h_{\mu\nu} =-2T_{\mu\nu}$ with the solution given by :
\begin{align}
\h_{\mu\nu}(x^\mu)|_{class}&= \frac{1}{\pi}\int d^4 x'\ \delta\big(\ (x-x')^2\ \big)\Theta(t-t')\ T_{\mu\nu}(x').\nn
\end{align}
The leading order solution around future null infinity is given by :
\begin{align}
\h_{\mu\nu}(u,r,\hat{x})|_{class}&= -\frac{1}{2\pi r}\int_{-\infty}^\infty du'\int d^2z'\ \frac{T^2_{\mu\nu}(\hat{x}',u')}{q.q'}.\nn
\end{align}
Above expression  is consistent with the fall offs mentioned in \eqref{h}. A key point about the perturbations is that $\p_u \h_{\mu\nu}^1=0$. This kills off a lot of terms that would have otherwise been present in \eqref{ephi}.
Finally we have : 
\begin{align}
\overset{\text{class}}{h_{rr}^1}(\hat{x})&=- \frac{1}{2\pi r}\int_{-\infty}^\infty du'\int d^2z'\ q.q'\ T^2_{uu}(\hat{x}',u').\label{hrrclassical}
\end{align}

\subsection{Quantum part}
Next we want to check if there is a part of $h_{rr}^1, A_r^1$ that has not been captured by the retarded propagator. Let us work in $r, u \rightarrow \infty$ limit ($u<r$) where the sources have died down and we can use the homogenous solution. We will use Herdegen-like  representation of $h_{\mu\nu}$. Herdegen representation\cite{herdegen} for photon is a way to write a generic homogenous solution for gauge field in Lorenz gauge in terms of free data $A_A^0$. Similarly, here we write a generic homogenous solution for metric field in De Donder gauge in terms of free data $C_{AB}$. ($C_{AB}=\lim_{ r\rightarrow \infty} \frac{h_{AB}(x)}{r}$). We have (See Appendix \ref{C} for details ) :

\begin{align}
 \overset{hom}{h_{\mu\nu}}(x) 
   &=- \frac{1}{(4\pi)}\int d^2z' \ (1+z'\bar{z}')^2\ \Big[ \epsilon_\mu^-\epsilon_\nu^-\ \dot{C}_{zz}(u=-x\cdot q',\hat{q}') +  \epsilon_\mu^+\epsilon_\nu^+\ \dot{C}_{\bar{z}\bar{z}}(u=-x\cdot q',\hat{q}')\ \Big], \label{herdegen}
\end{align}
$q'^\mu$ is defined according to \eqref{q}. From above expression it can be seen that $C_{zz}\sim \overset{\log}{C_{zz}} \log u$ gives rise to a $\frac{1}{r}$ term in $h_{\mu \nu}$. Let us find the $h_{rr}^1$ term by co-ordinate transformation. We will denote it with a 'quan' overtext. The expression for $h_{rr}^1$ term is thus given by :
 \begin{align}
\overset{quan}{h_{rr}^1}(x)\ &= \frac{1}{4\pi}\ \int  d^2z'\ (1+z'\bar{z}')^2  \frac{1}{q'.q}\ [\epsilon^-. q\ \epsilon^-. q\ \overset{\log}{C_{zz}}(\hat{x}') + \epsilon^+. q\ \epsilon^+. q\ \overset{\log}{C_{\bar{z}\bar{z}}}(\hat{x}')].\label{hrr quan}
\end{align}
Hence, we see that $\log u $ mode in $C_{AB}$ contributes to $h_{rr}^1$. We will eventually see that the existence of a $\log u $ mode is intimately tied to the leading soft theorem.

The behaviour of the free data around $u\rightarrow\pm\infty$ dictated by tree level soft theorems is :
%We discuss a purely quantum mode present in both metric and gauge fields that will eventually contribute to the charge. Let us  denote gravitational free data by $C_{AB(u,\hat{x})}$. ($C_{AB}=\lim_{ r\rightarrow \infty} \frac{h_{AB}(x)}{r}$).
\begin{align}
C_{AB}\ =\  D_{AB}^\pm(\hat{x})\ u^0\ +   ...\ , \ \  u \rightarrow \pm\infty. \label{uh}
\end{align}
Here, '...' denote any fall offs faster than power law fall off in u. Power law fall offs appear in above equation when we include the effect of long range forces. But these power law terms are not relevant our analysis. Just by Fourier transform one can quickly check that $D_{AB}^\pm$ is related to the leading soft factor or see \cite{log waves} for details. We will show that there is an additional term in \eqref{uh} :
\be
C_{AB}\ =\ \overset{\log}{C_{AB}}\ \log|u| + D_{AB}^\pm(\hat{x})\ u^0\ +   ...\ , \ \  \  u \rightarrow \pm\infty,
\ee
where, $\overset{\log}{C_{AB}}$ vanishes classically. An important point to note is that we are not introducing a new independent mode in the quantum system. $\overset{\log}{C_{AB}}$ is not arbitrary but is fixed in terms of the leading soft factor i.e. $D^\pm_{AB}$, hence the free data for classical system is sufficient to describe the quantized system as well. 

In \cite{1903.09133}, authors discussed the discontinuity in $\omega \tilde{A}_A$ as $\omega \rightarrow 0$ that is non trivial at quantum mechanical level. This discontinuity leads to a $\log | u| $ term in $A_A$.\footnote{It is interesting to note that this $\log u$ mode has appeared in equation (A.2) of \cite{prahar}.} We will discuss the gravitational analogue of this purely quantum $\log |u|$ mode. For scalars, $\omega \tilde{\phi}$ as $\omega \rightarrow 0$ is  trivial. Hence there is no $\log |u|$ term for scalars.

Let us consider $C^+_{zz}(u,\hat{x})$ that has only positive frequencies. We know that around $\omega\sim 0$, the behaviour of the radiative data is given by $\tilde{C}_{zz}^{+}\ = \ \frac{1}{\omega}\tilde{C}_{zz}^{+0}+...$. This low energy behaviour dictates the large-$u$ behaviour. Hence :
\begin{align}
C^+_{zz}(u,\hat{x})&= \frac{1}{2\pi }\int_0^\infty d\omega\ \big[\frac{1}{\omega}\tilde{C}_{zz}^{+0}(\hat{x})+...\big]\ e^{-i\omega u},\nn\\
&= \frac{1}{2\pi }\ \log(u^{-1})\ \tilde{C}^{+0}_{zz}(\hat{x}) + ...\ .
\end{align}
Simlarly for negative frequencies, we have :
\begin{align}
C^-_{zz}(u,\hat{x})&=- \frac{1}{2\pi }\ \log(u^{-1})\ \tilde{C}^{-0}_{zz}(\hat{x}) + ...\ .
\end{align}
Collecting the positve and negative frequency terms we get : 
\begin{align}
C_{zz}(u,\hat{x})& =\ -\frac{1}{2\pi} \big[\tilde{C}^{+0}_{zz}(\hat{x})-\tilde{C}^{-0}_{zz}(\hat{x})\big] \ \  \log |u| +...\ ,\nn\\
& =\ -\frac{1}{2\pi}\lim_{\omega\rightarrow 0^+} \ \big[\omega\tilde{C}^+_{zz}(\omega,\hat{x})+\omega\tilde{C}^-_{zz}(-\omega,\hat{x})\big] \ \  \log |u| +...\ .
\end{align}

Above expression tells us that the $\log u$ term is governed by the discontinuity in $\tilde{C}_{AB}$ as $\omega\rightarrow 0$. This term is absent in the classical theory. To find classical radiation, we use  retarded propagators. For such solutions, $\omega\tilde{C}_{zz}$ is continuous at $\omega=0$ \cite{log waves} and the coefficient of $\log |u|$ term vanishes. It is important to note that the coefficient does not carry any factor of $\hbar$. Hence, the $\log |u|$ coefficient does not go to 0 just by taking $\hbar\rightarrow 0$. The coefficient vanishes when we demand retarded boundary conditions. We will see that it is non-trivial quantum mechanically. This is because of the fact that when we quantise the $C_{AB}$ field, the positive frequencies involve annihilation operator while negative frequencies involve creation operator :
\be \tilde{C}_{zz}(\omega,\hat{x}) = \frac{-ic_{+}(\omega,\hat{x})}{2\pi(1+z\bar{z})^2} \ \ ...\ \ \omega >0,\ \ \ \tilde{C}_{zz}(\omega,\hat{x}) = \frac{ic^\dagger_{-}(-\omega,\hat{x})}{2\pi(1+z\bar{z})^2} \ \ ...\ \ \omega <0.\ee 
Thus, we get : 
\begin{align}
C_{zz}(u,\hat{x})& =\  \frac{i}{4\pi^2} \frac{1}{(1+z\bar{z})^2}\ \lim_{\omega\rightarrow 0}\ \omega[c_+(\omega,\hat{x})+c^\dagger_-(-\omega,\hat{x})] \ \  \log |u| +...\ .\label{Czz log}
\end{align}
Similarly, for $C_{\bar{z}\bar{z}}$ we have,
\begin{align}
 \overset{\log}{C_{\bar{z}\bar{z}}}(\hat{x}) &= \frac{i}{4\pi^2} \frac{1}{(1+z\bar{z})^2}\ \lim_{\omega\rightarrow 0}\ \omega[c_-(\omega,\hat{x})+c^\dagger_+(-\omega,\hat{x})].\label{Cbarz log}
\end{align}
We will see that above operators have non-trivial action when inserted in the expression for charge.
%
%
%
%Next, we want to find the correction to $h_{rr}^1$  due to the $\log u$ term. 
Substituting for $\overset{\log}{C}_{AB}$ in the expression of $\overset{quan}{h_{rr}^1}$ given in \eqref{hrr quan} :
 \begin{align}
\overset{quan}{h_{rr}^1}(x)\ &= \frac{1}{4\pi}\ \int  d^2z'\ (1+z'\bar{z}')^2  \frac{1}{q'.q}\ [\epsilon^-. q\ \epsilon^-. q\ \overset{\log}{C_{zz}}(\hat{x}') + \epsilon^+. q\ \epsilon^+. q\ \overset{\log}{C_{\bar{z}\bar{z}}}(\hat{x}')].\label{hrr quan1}
\end{align}
Next we need to do the sphere integral. We have relegated this calculation to Appendix \ref{BB} and we will quote the results here. The finite part of the integral is :
\begin{align}
<\text{out}|\overset{quan}{h_{rr}^1}(x)\ S\ |\text{in}>
&=-\frac{i}{2\pi^2}\ \sum_j ({q}.{p}_j)\ \log(q.{p}_j). \ \label{hrrhomo}
\end{align}
Next we repeat the calculation for gauge field and we get :
\begin{align}
<\text{out}|\overset{quan}{A_{r}^1}(\hat{x})\ S\ |\text{in}>  &=-\frac{i}{4\pi^2}\ \sum_j e_j\ \log(q.{p}_j). \ \label{Arhomo}
\end{align}
We note that above expressions can be obtained also by replacing the retarded propagators in section 6.1 with Feynman propagators. Finally, we have the complete expressions for  $h_{rr}^1$ and $A_r^1$. In the next section, we will use these expressions to evaluate the action of hard charge in the next section.

\section{The Ward identity}
The Ward identity for $S$ matrix for the 1-loop asymptotic charge can be written down as :
\begin{align}
\Big[\ Q^{\text{1-loop}}\ , \ S\  \Big] &= 0,\nn\\
\Rightarrow \Big(Q^{\text{soft}}_+\ S\ -\ S\ Q^{\text{soft}}_- \ \Big)&=-\Big(Q^{\text{hard}}_+\ S\ -\ S\ Q^{\text{hard}}_- \ \Big).\nn
\end{align}
Using \eqref{hard} and \eqref{hardp}, we get 
\begin{align}
\Big(Q^{\text{soft}}_+\ S\ -\ S\ Q^{\text{soft}}_- \ \Big)&
&=-\int d^2z'\ D_{z}'^2 V^z\ \gamma^{z\bar{z}}\ \Big({A^{0,1}_{\bar{z}}}(\hat{x}')\ S\ -\ S\ {B}^{\log}(\hat{x}') \ \Big).
\end{align}
Next we need to evaluate the action of above operators on a Fock state. From \eqref{1/ufull}, we get the expression of $A^{0,1}_{\bar{z}}$.
\begin{align}
A^{0,1}_{\bar{z}}(\hat{x})&=\ \frac{\sqrt{\gamma_{z\bar{z}}}}{4\pi } \int_{-\infty}^\infty du'\int { d^2z'}\  \frac{q^\mu\  \epsilon^\sigma _-}{q.q'}\ q'_{[\sigma}\p'_{q^\mu]}\ [- \ \frac{1}{2}h_{rr}^1(\hat{x}')j_u^2(\hat{x}')\ + 2 e^2A_r^1(\hat{x}')|\phi^1(\hat{x}')^2\ ]\nn\\
&-\frac{\sqrt{\gamma_{z\bar{z}}}}{8\pi } \ h^1_{rr}(\hat{x})\ \lim_{\omega\rightarrow 0}\omega a_-(\omega,\hat{x}).\label{A1u}
\end{align}
It is interesting to note that the first line resembles tree level subleading soft operator acting on $h_{rr}^1+A_r^1$. Similarly the second line is $h_{rr}^1$ times the leading soft operator.
The action of \eqref{A1u} on an outgoing Fock state can be easily evaluated. 
\begin{align}
<\text{out}|\ Q^{\text{hard}}_+&=\ <\text{out}|\ 4\pi \sum_{i\ \epsilon\ out}\ U^{\sigma\mu}(q_i)\   q_{i[\sigma}\p_{q_i^\mu]}\ [\ \frac{e_i}{2} h_{rr}^1(z_i) \ +\   e_i^2\frac{A_r^1(z_i)}{\omega_i}\ ]\nn\\
&-<\text{out}|\  \int d^2z'\ D_{\bar{z}}'^2 V^z(z')\ \frac{\sqrt{\gamma'_{z\bar{z}}}}{8\pi}\   \sum_i \frac{e_i\  \epsilon^-. \ p_i}{q'.p_i}\  h_{rr}^1(z')  .\label{H1}
\end{align}
where we have defined \be U^{\sigma\mu}(q_i)=\int d^2z'\ D_{z}'^2 V^z(z,z')\  \frac{\sqrt{\gamma'^{z\bar{z}}}}{16\pi^2}\  \frac{  \epsilon^\sigma _-q'^\mu}{q'.q_i},\label{U}\ee to make the expressions compact. Similarly we can use \eqref{Blog} to get the action of the past hard charge on an incoming state. Then we need to substitute for $h_{rr}^1$ and $A_r^1$.\\\\
\textbf{Classical part}\\
Let us first write down the purely electromagnetic term :
\begin{align}
&<\text{out}|\ \Big[Q^{\text{hard}}\ ,\ S\  \Big]_{class}\ |\text{in}>\ =- \sum_{i}e_i\ U^{\sigma\mu}(q_i)\ \  <\text{out}|\   \Big[ q_{i[\sigma}\p_{q_i^\mu]}\ \overset{\text{class}}{A_r^1}(\hat{x}_i)\ , \ S\ \Big] \ |\text{in}>.
\end{align}
Using \eqref{Arclassical}, we see that the classical part of $A_r^1$ is trivial and above term vanishes. This reflects the absence of classical $\log \omega$ term in soft electromagnetic radiation (in absence of gravitational coupling)\cite{1808.03288}.

Next we substitute the classical part of $h_{rr}^1$ using \eqref{hrrclassical} to get :
\begin{align}
&<\text{out}|\ \Big[Q^{\text{hard}}\ ,\ S\  \Big]_{class}\ |\text{in}>\nn\\
&=- \sum_{i,j;\eta_i\eta_j=1}e_i\ U^{\sigma\mu}(q_i)\ \     q_{i[\sigma}\p_{q_i^\mu]}\ (p_j.q_i)\ \ \mathcal{M}_n+\sum_i e_i\  q_{i\sigma}\ U^{\sigma \mu}\ \sum_{j;\eta_j=1}  p_{j\mu}\ \ \mathcal{M}_n
%&=-\sum_{i,j;\eta_i\eta_j=1} e_i\  U^{\sigma\mu}(q_i) \  (p_{j\mu}q_{i\sigma}\ - \ p_{j\sigma}q_{i\mu}\ )\ \ \mathcal{M}_n+\sum_i e_i\  q_{i\sigma}\ U^{\sigma \mu}\ \sum_{j;\eta_j=1}  p_{j\mu}\ \ \mathcal{M}_n.
 \label{h1}
\end{align}
Here, we have $\mathcal{M}_n = <\text{out}|\  S \ |\text{in}>.$
\\\\
\textbf{Quantum part}\\
Next we will derive the quantum pieces. Using \eqref{Arhomo} for $A_r^1$, the purely electromagnetic term turns out to be : 
\begin{align}
& <\text{out}|\ \Big[Q^{\text{hard}}\ ,\ S\  \Big]_{quan}\ |\text{in}>\nn\\
&=-\frac{i}{\pi}\ \sum_{i,j;i\neq j}\frac{e^2_ie_j}{\omega_i}\ U^{\sigma\mu}(q_i)\ \     q_{i[\sigma}\p_{q_i^\mu]}\  \log[\frac{2(p_j.q_i)}{m_j^2}]\ \big]\ \ \mathcal{M}_n.
\end{align}
The term that depends on $m_j$ is the divergent piece in $A_r^1$. It is killed by the derivative operator $q_{i[\sigma}\p_{q_i^\mu]}$. Next we will use the quantum part of $h_{rr}^1$ from \eqref{hrrhomo}. Substituting in the expression for hard charge we get :\\
\begin{align}
& <\text{out}|\ \Big[Q^{\text{hard}}\ ,\ S\  \Big]_{quan}\ |\text{in}>\nn\\
&=-\frac{i}{\pi}\ \sum_{i,j;i\neq j} e_i\ U^{\sigma\mu}(q_i)\ \     q_{i[\sigma}\p_{q_i^\mu]}\ \  p_j.q_i\ \log[\frac{2(p_j.q_i)}{m_j^2}] \ \mathcal{M}_n\nn\\
&+i\int d^2z'\ D_{\bar{z}}'^2 V^z(z')\ \frac{\sqrt{\gamma'_{z\bar{z}}}}{16\pi^3}\ \sum_i \frac{e_i\  \epsilon^-. \ p_i}{q'.p_i} \sum_{j\ } q'.p_j \log [\frac{2(p_j.q')}{m_j^2}] \ \ \mathcal{M}_n.
\end{align}
 The two divergent terms in above expression cancel each other. Thus, we get a finite action of the charge. The full quantum term is given by :
\begin{align}
& <\text{out}|\ \Big[Q^{\text{hard}}\ ,\ S\  \Big]_{quan}\ |\text{in}>\nn\\
&=-\frac{i}{\pi} \sum_{i,j;i\neq j} e_i\  U^{\sigma\mu}(q_i)\ \     q_{i[\sigma}\p_{q_i^\mu]}\  \big[p_j.q_i\ \log(p_j.q_i)+\frac{e_ie_j}{\omega_i} \log(p_j.q_i)\ \big]+ \frac{i}{\pi}\sum_i e_i\  q_{i\sigma} \ \sum_j \tilde{U}_j^{\sigma \mu}\ p_{j\mu} \ \mathcal{M}_n.\ \label{h2}
\end{align}
here we have defined \be \tilde{U}_j^{\sigma\mu}(q_i)=\int d^2z'\ D_{z}'^2 V^z(z,z')\  \frac{\sqrt{\gamma'^{z\bar{z}}}}{16\pi^2}\  \frac{  \epsilon^\sigma _-q'^\mu}{q'.q_i} \log (q'.p_j).\label{Utilde}\ee

Collecting together \eqref{h1} and \eqref{h2} we get the complete action of the hard charge and we can write down the Ward identity. Thus, the $S$-matrix needs to satisfy following Ward identity for a generic $V^z$ that lives on $S^2$ : 
\begin{align}
\Big[Q^{\text{soft}}(V^z)\ ,\ S\  \Big]
&=\ -\ C_{\text{hard}}(V^z)\ \  S. \label{ward}
\end{align}
$Q^{\text{soft}}(V)$ defined in \eqref{soft1} inserts soft modes of photon. We have :
\begin{align}
C_{\text{hard}}(V^z)
&= \ \sum_i e_i\  q_{i\sigma}\ U^{\sigma \mu}\ \sum_{j;\eta_j=1}  p_{j\mu}\ -\ \sum_{i,j;\eta_i\eta_j=1}e_i\   U^{\sigma\mu}(q_i) \  (p_{j\mu}q_{i\sigma}\ - \ p_{j\sigma}q_{i\mu}\ )\nn\\
&-\frac{i}{\pi} \sum_{i,j;i\neq j} e_i\  U^{\sigma\mu}(q_i)\ \Big[\frac{e_ie_j}{q_j.q_i} (q_{j\mu}q_{i\sigma}\ - \ q_{j\sigma}q_{i\mu}\ )+(p_{j\mu}q_{i\sigma}\ - \ p_{j\sigma}q_{i\mu}\ )\ \log(-p_j.p_i)\ \Big]\footnotemark\nn\\
&  +\  \frac{i}{\pi}\sum_i e_i\  q_{i\sigma} \ \sum_j \tilde{U}_j^{\sigma \mu}\ p_{j\mu}\ .  \label{Co}
\end{align}
\footnotetext{The first term in \eqref{h1} produces a term that vanishes due to conservation of momenta. This is the term discussed in footnote 2.}
Dependence on $V^z$ is via the $U$'s defined in \eqref{U} and \eqref{Utilde}. Ward identity involving $V^{\bar{z}}$ can be written down similarly.
\subsection{The Sahoo-Sen soft theorem}
Let us derive the Sahoo-Sen soft theorem from above Ward identity.  To derive negative helicity soft theorem we choose  \cite{1903.09133} :
\be 
V^z(z,z') = {\sqrt{2}}(1+z'\bar{z'})\frac{z-z'}{\bar{z}-\bar{z}'},\ \ V^{\bar{z}}=0 \label{V}.
\ee
Performing the sphere ($z',\bar{z}'$) integral in \eqref{soft1}, we get :
\begin{align}
Q_+^{\text{soft}}
&=-i\lim_{\omega\rightarrow 0}  \omega\ \partial_\omega^2\ \omega\  \ a_{-}(\omega,\hat{x}) .\label{soft}
\end{align}
Next we will use \eqref{V} in the expression for hard charge \eqref{Co}. The sphere integral in the expression for $U$ \eqref{U} and for $\tilde{U}$ in \eqref{Utilde} can be done easily. We get :
\begin{align}
C_{\text{hard}}
&=\frac{1}{4\pi} \sum_i e_i\frac{\epsilon.p_i}{p_i.k}\ \sum_{j;\eta_j=1}k.p_j\ \ -\ \frac{1}{4\pi}\sum_{\substack{i,j;i\neq j \\ \eta_i\eta_j =1}} e_i\frac{\epsilon_\mu k_\rho}{ p_i.k}(p_j^\rho p_i^\mu-p_i^\rho p_j^\mu)\nn\\
& -\frac{i}{4\pi^2}\sum_{i,j;i\neq j}e_i\frac{\epsilon_\mu k_\rho}{ p_i.k}\  \Big[ \frac{e_ie_j}{p_i.p_j}(p_j^\rho p_i^\mu-p_i^\rho p_j^\mu)+(p_j^\rho p_i^\mu-p_i^\rho p_j^\mu)\log[p_i.p_j]\ \Big]\nn\\
& +\ \frac{i}{4\pi^2}\sum_i e_i\frac{\epsilon.p_i}{ p_i.k}\sum_{j} k.p_j\ \log\ {{p_j.q}}\ .
\end{align}
 So, the Ward identity can be recast as : 
\begin{align}
&\lim_{\omega\rightarrow 0}  \omega\ \partial_\omega^2\ \omega \ \mathcal{M}_{n+1}\nn\\
 &=\Bigg[-\frac{i}{4\pi} \sum_i e_i\frac{\epsilon.p_i}{p_i.k}\ \sum_{j;\eta_j=1}k.p_j\ \ +\ \frac{i}{4\pi}\sum_{\substack{i,j;i\neq j \\ \eta_i\eta_j =1}} e_i\frac{\epsilon_\mu k_\rho}{ p_i.k}(p_j^\rho p_i^\mu-p_i^\rho p_j^\mu)\nn\\
& -\frac{1}{4\pi^2}\sum_{i,j;i\neq j}e_i\frac{\epsilon_\mu k_\rho}{ p_i.k}\  \Big[ \frac{e_ie_j\ }{p_i.p_j}(p_j^\rho p_i^\mu-p_i^\rho p_j^\mu)+(p_j^\rho p_i^\mu-p_i^\rho p_j^\mu)\log[p_i.p_j]\ \Big]\nn\\
& +\ \frac{1}{4\pi^2}\sum_i e_i\frac{\epsilon.p_i}{ p_i.k}\sum_{j} k.p_j\ \log\ {{p_j.q}}\ \Bigg]\ \mathcal{M}_n.
\end{align}

This is exactly the Sahoo-Sen soft theorem. In this analysis we have derived the soft theorem from the Ward identity. The Ward identity (with $V^{\bar{z}}=0$) can be derived from the soft theorem by multiplying both sides of the statement of soft theorem with $\int d^2z\ D_{\bar{z}}^2 V^z(z)\ \frac{\sqrt{\gamma_{z\bar{z}}}}{16\pi^2}\ $. Thus, we can conclude that the Ward identity \eqref{ward} is exactly equivalent to the Sahoo-Sen soft photon theorem \eqref{softthm2}. 

Soft theorems are expected to be related to asymptotic symmetries. It is well known that QED amplitudes exhibit leading soft theorem that is equivalent to Ward identity of large U(1) gauge transformations. In this paper we studied this equivalence for loop level subleading soft theorem. This study was initiated in \cite{1903.09133} for the case of massive scalar QED. We showed that the Sahoo-Sen soft photon theorem for massless scalar QED coupled to gravity is equivalent to the conservation law in \eqref{cons1}. It would be interesting to understand the symmetry underlying this conservation law. % In \cite{log memory}, the authors discussed the direct experimental consequence of above loop level soft graviton theorem.  Interestingly, they have predicted a new tail to the well known memory effect \cite{mem1,mem2,mem3}.
\section{Acknowledgements}
I am deeply thankful to Nabamita Banerjee, Miguel Campiglia and Alok Laddha for many helpful discussions and suggestions. I thank Chennai Mathematical Institute where part of
this work was done. I am grateful to the people of India for their 
support to theoretical sciences.

\appendix

\section{Calculating the $1/u$ mode in $A_A^0$}\label{AA}
Dressing of scalars under long range forces lead to logarithmic modes in the current :
$$j_A\ =\\ j_A^{\log}\ \frac{\log r}{r^2}+ \frac{{j}^2_A}{r^2} \ + ...\ .$$
We also have :
$$j_r = j_r^{\log}\frac{\log r}{r^4}+\frac{j_r^4}{r^4}+...\ \ , j_u\ =\ \frac{{j}^2_u}{r^2} \ + \ j_u^{\log}\ \frac{\log r}{r^3}+...\ .$$
For the Cartesian components of the U(1) current we have :
\be j_\mu\ =\ \frac{j^2_\mu}{r^2} \ + \ {j^{\log}_\mu}\ \frac{\log r}{r^3}+...\ .\ee
We will substitute above current source in :
\begin{align}
A_\sigma(x)= \frac{1}{2\pi }\int d^4x'\ \delta(\ (x-x')^2)\  \Theta(t-t')\ j_\sigma(x').
\end{align}
 Let us take the limit $r\rightarrow \infty$ keeping $u$ finite :
\begin{align}
&A_\sigma(u,r,\hat{x})\nn\\
&= \frac{1}{4\pi r}\int_{-\infty}^\infty du'\int_0^\infty dr'  \int_{S^2}\frac{ d^2z'}{-q.q'}\ \delta(r'+\frac{u-u'}{q.q'})\  \Big[ {j^2_\sigma}(u',z')\ + {j^{\log}_\sigma}(u',z')\ \frac{\log r'}{r'}+{j^3_\sigma}(u',z')\ \frac{1}{r'}+...\Big] ,\nn\\
&= \frac{1}{4\pi r}\int_{-\infty}^\infty du'\int _{S^2} d^2z'\ \Bigg[\frac{{j^2_\sigma}(u',z')}{-q.q'}\ +{j^{\log}_\sigma}(u',z')\  \frac{\log(u-u')}{u-u'} +\frac{[{j^3_\sigma}(u',z')-{j^{\log}_\sigma}(u',z')\  \log (-q.q')]}{u-u'}\Bigg].\label{11}
\end{align}
We are interested in studying the u-behaviour in $u\rightarrow \infty$ limit. In \eqref{11}, the $j^2_\sigma$ term contributes to $u^0$ term as $u\rightarrow \infty$. The next dominant fall off in $u\rightarrow \infty$ limit is $\frac{\log u}{u}$. It comes from the region $u'<<u$. Thus, we have : 
\begin{align}
A_\sigma(u,r,\hat{x})
&= \frac{1}{4\pi r}\int_{-\infty}^\infty du'\int_{S^2}  d^2z'\ \Bigg[\frac{{j^2_\sigma}(u',z')}{-q.q'}\ +{j^{\log}_\sigma}(u',z')\  \frac{\log u}{u} +...\Bigg].\label{10}
\end{align}First we rewrite the coefficient in retarded co-ordinates (Recalling that $q^{\mu} = (1,\hat{x})$.) : 
\begin{align}
{j^{\log}_\sigma} 
&= -q_\sigma j_u^{\log}+\gamma^{AB}\p_B{q_\sigma}\ {j^{\log}_A}.\label{jlog}
\end{align}
$ j_u^{\log}$ can be eliminated using the conservation equation of current :
\be
 \ {j^{\log}_u}=\p_u{j^{\log}_r}-D^A{j^{\log}_A}.\label{julog}\ee
Substituting in the expression for ${j^{\log}_\sigma} $ :
\begin{align}
 {j^{\log}_\sigma} 
&= -q_\sigma\p_u{j^{\log}_r} +D^A[{q_\sigma}\ {j^{\log}_A}].\label{jsigma}
\end{align}
Thus, ${j^{\log}_\sigma}$ is a total derivative. When \eqref{jsigma} is substituted in \eqref{10}, the $D^A[{q_\sigma}\ {j^{\log}_A}]$ term vanishes trivially due to sphere integral. Using the logarithmic fall off of the gauge field : $ A_r=\frac{A_r^1}{r}+A_r^{\log}\frac{\log r}{r^2}+...$ in the expression of U(1) current we get 
\be j_r^{\log} = -2 e^2 A_r^{\log} |\phi^1|^2. \label{jr}\ee
 Using \eqref{jr} let us study the behaviour of $j_r^{\log}$ as $|u|\rightarrow \infty$. Following the logic of \cite{1605.09677}, we know that $\phi \sim \frac{1}{u^{1+\epsilon}}$ as $|u|\rightarrow \infty$. Now, let us find the $u$-fall off of $A_r^{\log}$. Using the gauge condition we have : $\p_uA_r^{\log} = -A_u^{\log}$. Then $A_u^{\log}$ can be related to the current by Maxwell's equation :  $2\p_uA_u^{\log} = j_u^2$. Hence, $A_r^{\log}$ can have a $\mathcal{O}(u)$ term as $|u|\rightarrow \infty$. Using these u-fall offs in the expression \eqref{jr} we get $j_r^{\log}\rightarrow 0$ as $|u|\rightarrow \infty$. Thus, the first term in \eqref{jsigma} also gives a vanishing contribution. Hence the coefficient of $\frac{\log u}{u}$ vanishes.  \\
The next term falls off as $1/u$ and this is the term that is relevant for loop level charge. Let us rewrite the $1/u$-term in a nice form. To start with, we have : 
\begin{align}
A_\sigma(u,r,\hat{x})&= \frac{1}{4\pi ru}\int_{-\infty}^\infty du'\int_{S^2} d^2z'\ \Big[{j^3_\sigma}(z')-{j^{\log}_\sigma}(z')\  \log (-q.q')\Big].\nn
\end{align}
We manipulate ${j^3_\sigma}$ in similar fashion : 
\begin{align}
 {j^3_\sigma} 
&= -q_\sigma j_u^{3} +\gamma^{AB}\p_B{q_\sigma}\ j_A^2\ =\  -q_\sigma\p_uj^4_r -q_\sigma j^{\log}_u+D^A[\ {q_\sigma}\ j_A^2],
\end{align}
and we get :
\begin{align}
A_\sigma(u,r,\hat{x})&= \frac{1}{4\pi ru}\int_{-\infty}^\infty du'\int_{S^2}  d^2z'\ \Big[- q'_\sigma[{j^{\log}_u}+\p'_uj_r^4]+\big[q'_\sigma\p_u{j^{\log}_r} -D'^A[{q'_\sigma}\ {j^{\log}_A}]\big]\  \log (-q.q')\Big].\nn\end{align}
%\footnote{$$\f{\p z^A}{\p x^i}=\f{\gamma^{AB}}{r}\ \f{\p q_i}{\p z^B}$$}
We again substitute for ${j^{\log}_u}$ using \eqref{julog}. Upto total sphere derivative terms, above expression can be rewritten as :
\begin{align}
A_\sigma(u,r,\hat{x})
&= \frac{1}{4\pi ru}\int_{-\infty}^\infty du'\int_{S^2}  d^2z'\ \Big[\ q'_\sigma[D'^A{j^{\log}_A}+{j^{\log}_A}\ D'^{A} \log (-q.q') ]-q'_\sigma\ \p'_u [{j^{\log}_r}+j_r^4-{j^{\log}_r}\log (-q.q')]\ \Big].\nn \label{pujr}\end{align}
The last term drops out as $j_r^4\rightarrow 0$ as $|u|\rightarrow \infty$ (proved in \cite{1605.09677}) and we have already checked that $j_r^{\log}\rightarrow 0$ as $|u|\rightarrow \infty$. We can rewrite the first term as 
\begin{align}
A_\sigma(u,r,\hat{x})
&= \frac{1}{4\pi ru}\int_{-\infty}^\infty du'\int_{S^2}  d^2z'\ q^\mu\ \frac{q'_{[\sigma}D'^A q'_{\mu]}}{q.q'}\ {j^{\log}_A},
\end{align}
where, $q_{[\mu}\ D^A q_{\nu]}= q_\mu\ D^A q_\nu \ - \ q_\nu\ D^A q_\mu.$ Finally we perform a co-ordinate transformation (using \eqref{q}) to get :
\begin{align}
A^0_{\bar{z}}(u,\hat{x})
&= \frac{1}{4\pi u}\frac{\sqrt{2}}{1+z\bar{z}}\int_{-\infty}^\infty du'\int _{S^2} d^2z'\ \frac{\epsilon_-^\sigma q^\mu}{q.q'}\ q'_{[\sigma}D'^A q'_{\mu]}\  {j^{\log}_A}.\label{Ascalar1}
\end{align}
\textbf{At past null infinity} :\\
We recall that the charge at past is defined in terms of following mode \eqref{cons} : $ \frac{\log r}{r^2}F_{rA}(\hat{x})|_{\mathcal{I}^-_+}$. To study this mode first we will expand Maxwell's equations in large r at finite v and take $v\rightarrow\infty$ limit in the solution. Around $\mathcal{I}^-$, the gauge field equation $\Box A_\mu = -j_\mu$  is :
$$ [2\p_v\p_r + \frac{2}{r}\p_v + \p_r^2 +\frac{D^2}{r^2}]\ A_{\sigma} = -j_\sigma.$$
Using the asymptotic expansion for current :
$$ j_\mu = j^2_\mu\frac{1}{r^2}+j^{\log}_\mu\frac{\log r}{r^3}+...\ ,$$
we get :
$$ A_\mu =A^{\log 1}_\mu\frac{\log r}{r}+A^1_\mu\frac{1}{r}+A^{\log 2}_\mu\frac{\log r}{r^2}+...\ .$$
The coefficients satisfy :
\begin{align}
2\p_vA^{\log 1}_\sigma &= -j_\sigma^2,\nn\\
-2\p_vA^{\log 2}_\sigma +(D^2+2)\ A^{\log 1}_\sigma &= -j_\sigma^{\log}.\nn
\end{align}
The $\frac{\log r}{r^2}$ term in $F_{rA}$ comes from $A^{\log 2}_\sigma$. $A^{\log 1 }_\sigma$ is $\mathcal{O}(e)$ term and contributes an $\mathcal{O}(v)$ term at $v\rightarrow -\infty$. So, we ignore it henceforth :
\begin{align}
A^{\log 2}_\sigma(x)&= \frac{1}{2} \int^v_{-\infty} dv'  \overset{\log}{ j_\sigma}(v',\hat{x}) +\mathcal{O}(e)\ .
\end{align}
In above solution, we have chosen the integration constant such that $A^{\log 2}_\sigma \rightarrow 0$ as $v\rightarrow -\infty$. With a co-ordinate transformation, we get : 
\be F_{rz}^{\log 0}|_{\mathcal{I}^-_+}=-\frac{1}{2}\p_z q^\mu  \int _{-\infty}^{\infty}dv'  \overset{\log}{ j_\mu}(v',\hat{x}).\label{Flogr}\ee
Substituting for $\overset{\log}{ j_\sigma}$ we get :
\begin{align}
F_{rz}^{\log 0}|_{\mathcal{I}^-_+}&=- \frac{1}{2} \p_z q^\mu \int_{-\infty}^{\infty} dv'  D^A[q_\mu\ \overset{\log}{j_A}](v',\hat{x}).\nn
\end{align}
This can be rewritten as :
\begin{align}
F_{rz}^{\log 0}|_{\mathcal{I}^-_+}
&=- \frac{1}{2} \int _{-\infty}^{\infty}dv'   \overset{\log}{j_z}(v',\hat{x}),\nn\\
&= -\frac{1}{2} \int _{-\infty}^{\infty}dv' d^2z'\ \delta^2(\hat{x}-\hat{x}') j_{z}(v',\hat{x}'),\nn\\
&=- \frac{1}{4\pi}\ \int d^2z'\ D_z^2\ \Big[ \frac{q^\nu\p_{\bar{z}} q^\mu}{q.q'}\ q'_{[\mu}D^A q'_{\nu]}\ \overset{\log}{j_A}(\hat{x}')\Big].\nn
\end{align}
Above expression can be rewritten as :
\begin{align}
F_{rz}^{\log 0}|_{\mathcal{I}^-_+}
&= \frac{1}{4\pi} \int_{-\infty}^\infty dv'\int_{S^2} d^2z'\ \gamma^{z\bar{z}}\ D_z^2\ \Big[   \frac{q^\nu\p_{\bar{z}} q^\mu}{q.q'}\ q'_{[\mu}D^A q'_{\nu]}\ {j^{\log}_A}(v',-\hat{x}')\Big].\label{pastF}
\end{align}

\section{Quantum modes in $h_{rr}^1$ and $A_r^1$}\label{BB}
Let us start with the expression for $\overset{quan}{h_{rr}^1}$ given in \eqref{hrr quan1} :
 \begin{align}
\overset{quan}{h_{rr}^1}(x)\ &= \frac{1}{4\pi}\ \int  d^2z'\ (1+z'\bar{z}')^2  \frac{1}{q'.q}\ [\epsilon^-. q\ \epsilon^-. q\ \overset{\log}{C_{zz}}(\hat{x}') + \epsilon^+. q\ \epsilon^+. q\ \overset{\log}{C_{\bar{z}\bar{z}}}(\hat{x}')].
\end{align}
We will use the leading soft theorem to evaluate action of \eqref{Czz log} and \eqref{Cbarz log}. So, action of $h_{rr}^1$ on a generic state is given by :
\begin{align}
&<\text{out}|\overset{quan}{h_{rr}^1}(x)\ S\ |\text{in}>   \nn\\
&=i<\text{out}|\int \frac{d^2z'}{16\pi^3}\  \Big[ \frac{\epsilon^-\cdot q\ \epsilon^-\cdot q}{q'.q}\ \sum_j \frac{\epsilon^+\cdot p_j\ \epsilon^+\cdot p_j}{q'.p_j}\ + \ \frac{\epsilon^+\cdot q\ \epsilon^+\cdot q}{q'.q}\ \sum_j  \frac{\epsilon^-\cdot p_j\ \epsilon^-\cdot p_j}{q'.p_j}\Big]\ S\ |\text{in}>.\label{hrraction}
\end{align}
Using completeness relations for polarisation tensors :
\begin{align}
&  \frac{\epsilon^-\cdot q\ \epsilon^-\cdot q}{q'.q}\ \sum_j \frac{\epsilon^+\cdot p_j\ \epsilon^+\cdot p_j}{q'.p_j}\ + \ \frac{\epsilon^+\cdot q\ \epsilon^+\cdot q}{q'.q}\ \sum_j \frac{\epsilon^-\cdot p_j\ \epsilon^-\cdot p_j}{q'.p_j}= \sum_j \frac{2( q.p_j)^2}{q.q'\ \ q'.p_j}.
\end{align}
Thus, in \eqref{hrraction}, we need to do following integral : 
\begin{align}
I=\int d^2z'\ \frac{1}{q.q'\ \ q'.p_j}
&=\int d^2z'\ \int_0^1 dx\ \frac{1}{[\hat{q}'.(x\hat{q} + (1-x)\omega_j\hat{q}_j)-x-(1-x)\omega_j]^{2}},\nn\\
&=2\pi \int_0^1 dx\ \frac{1}{[x(1-x)\omega_j(1-\hat{q}_j.\hat{q})]}.\label{I}
\end{align}
But $I$ is divergent. These are collinear divergences that appear as we are dealing with massless particles. We will see that the diverging terms cancel and the charge is finite. Let us regulate the integral by introducing a regulator $m_j$ by making $p_j$ massive. Repeating previous steps for a massive $p_j$, we get :
\begin{align}
I&= \frac{4\pi}{q.p_j} \int_0^1 dx\ \frac{1}{[2x(1-x)+\frac{m_j^2}{q.p_j}(1-x)^2]}.\nn
\end{align}
Thus, $I$ still has divergence coming from $x=1$. But we will see that $x=1$ term vanishes due to conservation of momentum.
\begin{align}
I&= \frac{4\pi}{q.p_j}\ \frac{m^2_j-2q.p_j}{(-2q.p_j)}\ \Big[ \lim_{x\rightarrow 1}\log(1-x)+\log[\frac{m^2_j}{m_j^2-2q.p_j}]\ \Big]
\end{align}
Let us study above expression in the limit when the regulator is taken to 0 :
\begin{align}
\lim_{m_j\rightarrow 0}I
&=\ \frac{4\pi}{q.p_j}\ \Big[ \lim_{x\rightarrow 1}\log(1-x)+ \log[{{m^2_j}}]\ - \log[2q.p_j]\ +...\ \Big] .
\end{align}
Here, '...' denote terms that vanish when regulator is set to 0.
The infinite piece is as follows :
\begin{align}
<\text{out}|\overset{quan}{h_{rr}^1}(x)\ S\ |\text{in}> |_{inf}  &= \frac{i}{2\pi^2}\sum_j ({q}.{p}_j) \Big[ \lim_{x\rightarrow 1}\log(1-x)+ \log[{-\frac{m^2_j}{2}}]\ \Big]\nn\\
&= \frac{i}{2\pi^2}\sum_j ({q}.{p}_j)  \log[{\frac{m^2_j}{2}}].\ 
\end{align}
Here, the first piece vanishes due to conservation of momenta. We could have regulated the $x=1$ divergence right from the beginning by introducing a mass for the null vector $q^\mu$\footnote{The corresponding integral is same as the integral in equation (5.27) of \cite{1808.03288} and can be evaluated accordingly.} and gotten the same result for $I$. The finite piece is :
\begin{align}
<\text{out}|\overset{quan}{h_{rr}^1}(x)\ S\ |\text{in}>
&=-\frac{i}{2\pi^2}\ \sum_j ({q}.{p}_j)\ \log(q.{p}_j). \ 
\end{align}
Next we will repeat the calculation for gauge field. To start with, we have \cite{1903.09133} : 
 \begin{align}
\overset{quan}{A_{\mu}}(x)\ &= \frac{1}{2\pi}\  \int  d^2z'\frac{(1+|z'|^2)}{\sqrt{2}}  \frac{1}{q'.x}\ [\varepsilon^-_{\mu}\ \overset{\log}{A_{z}} + \varepsilon^+_{\mu}\ \overset{\log}{A_{\bar{z}}}],\label{Ar quan}
\end{align}
where,
\begin{align}
 \overset{\log}{A_{\bar{z}}}(\hat{x}') &= \frac{i}{8\pi^2} \frac{\sqrt{2}}{(1+|z'|^2)}\ \lim_{\omega\rightarrow 0}\ \omega[a_-(\omega,\hat{x}')+a^\dagger_+(-\omega,\hat{x}')],\nn\\
 \overset{\log}{A_{z}}(\hat{x}') &= \frac{i}{8\pi^2} \frac{\sqrt{2}}{(1+|z'|^2)}\ \lim_{\omega\rightarrow 0}\ \omega[a_+(\omega,\hat{x}')+a^\dagger_-(-\omega,\hat{x}')].
\end{align}
We extract out the $1/r$-term :
 \begin{align}
\overset{quan}{A_{r}^1}(\hat{x})\ &= \frac{1}{2\pi}\ q^{\mu} \int  d^2z'\frac{(1+|z'|^2)}{\sqrt{2}}  \frac{1}{q'.q}\ [\varepsilon^-_{\mu}\ \overset{\log}{A_{z}} + \varepsilon^+_{\mu}\ \overset{\log}{A_{\bar{z}}}].\label{Ar quan}
\end{align}
The action of $A_r^1$ can be evaluated on a generic out state :
\begin{align}
 <\text{out}|\overset{quan}{A_{r}^1}(\hat{x})\ S\ |\text{in}>  &=i<\text{out}|\int \frac{d^2z'}{16\pi^3} \ \Big[ \frac{ \varepsilon^-\cdot q}{q'.q}\ \sum_j e_j\ \frac{\varepsilon^+\cdot p_j}{q'.p_j}\ + \ \frac{ \varepsilon^+\cdot q}{q'.q}\ \sum_j e_j \frac{ \varepsilon^-\cdot p_j}{q'.p_j}\Big]\ S\ |\text{in}>,\nn\\
&=\frac{i}{16\pi^3} <\text{out}|\int d^2z'\ \sum_j e_j \frac{p_j\cdot q}{q'.q\ q'.p_j}\ S\ |\text{in}>.
\end{align}
Above integral can be calculated similar to the earlier one. The infinite piece is a constant :
\begin{align}
<\text{out}|\overset{quan}{A_{r}^1}(x)\ S\ |\text{in}> |_{inf}  &=\frac{i}{4\pi^2}\ \log[{\frac{m^2_j}{2}}].\ \label{Ainf}
\end{align}
 We have for the finite part :
\begin{align}
<\text{out}|\overset{quan}{A_{r}^1}(\hat{x})\ S\ |\text{in}>  &=-\frac{i}{4\pi^2}\ \sum_j e_j\ \log(q.{p}_j). \ 
\end{align}

\section{Maxwell's equations in presence of gravity}\label{A}
In this section we write down Maxwell's equations in presence of gravitational fluctuations given by \eqref{h}.\\
Let us study the $\nabla^\mu F_{u\mu} = j_u$ equation. Expanding the equation around $r\rightarrow \infty$, at $\mathcal{O}(\frac{1}{r^2})$ we get :
\begin{equation}
\partial_u\overset{2}{F}_{ru} + \partial_uD^BA_B^0-j_u^2=- \gamma^{CB}\ h^0_{Cr}\ \partial_u F^0_{uB}.\label{u comp}
\end{equation}
In the equation $\nabla^\mu F_{A\mu} = 0$, there appears a gravity correction even at $\mathcal{O}(\frac{1}{r})$ :
\begin{align}
\partial_u\overset{1}{F}_{rA}-h^1_{rr}\ \partial_u\overset{0}{F}_{Au}&=0.
\end{align}
This implies $\log r $ dressing of $A_A$ that has also been derived in \eqref{logr}. 
$\nabla^\mu F_{A\mu} = 0$ at $\mathcal{O}(\frac{1}{r^2})$ gives :
\begin{align}
&2\partial_u\overset{2}{F}_{rA}-\partial_A\overset{2}{F}_{ru} + D^BF_{AB}^0-j_A^2\nn\\
&=h^1_{rr}\ \partial_u\overset{1}{F}_{Au}+h^2_{rr}\ \partial_u\overset{0}{F}_{Au}- \gamma^{CB}\ h^0_{Cr}\ \partial_u F^0_{AB}
- \gamma^{CB}\ h^0_{Cr}\ D_B F^0_{Au}-\gamma^{BC}h^{-1}_{AB}F^0_{uC}\nn\\
&+\p_uh^2_{rr}\ \overset{0}{F}_{Au}+\frac{1}{2}h^1_{rr}\ \overset{0}{F}_{Au}-\frac{1}{2}\gamma^{BC}h^{-1}_{BC}F^0_{Au}-{\gamma^{BC}}D_Bh^0_{Ar}\ F^0_{uC}-D^Bh^0_{Br}\ F^0_{Au}
-2h^{1\ ur}\ F^0_{Au}.\label{A comp}
\end{align}
We use above equation to substitute for $\p_u\overset{2}{F}_{rA}$ in \eqref{Q} i.e. in 
\begin{align}
Q_+^{\text{soft}}&=-\int du\  d^2z'\  V^A  \partial_u\ \big[u^2 \partial^2_u\overset{2}{F}_{rA} \big],
\end{align}
 and we get \eqref{Qs} i.e.
\begin{align}
Q_+^{\text{soft}}&=-\frac{1}{2}\int du\  d^2z'\  V^A  \partial_u\ \big[u^2 \partial_u[\partial_A\overset{2}{F}_{ru} - D^BF_{AB}^0+j_A^2]\ \big]+...,
\end{align}
 where "..." refers to the gravity corrections that come from RHS of \eqref{A comp} and \eqref{u comp}. We will analyse them one by one. Out of the metric components appearing in Maxwell's equations only $h_{rr}^2$ and $h_{AB}^{-1}$ depend on $u$, rest of them are $u$-independent. This simplifies the analysis for most of the terms.\\
\textbf{Term $h^1_{rr}\ \partial_u{F}^1_{Au}$}
\begin{align}
Q_1^{\text{cor}}&=-\frac{1}{2}\int du\ d^2z\  V^z  \partial_u\ \Big[u^2 \partial_u[h^1_{rr}\ \partial_u\overset{1}{F}_{uz}]\Big]\ +\ z\leftrightarrow \bar{z}.\nn
\end{align}
Using Bianchi identities we can simplify above expression to  :
\begin{align}
Q_1^{\text{cor}}&=-\frac{1}{2}\int d^2z\  V^z\int du\ h^1_{rr}\  \partial_u\ \Big[u^2 \partial^2_uD_z^2A^0_{\bar{z}}\Big] +\ z\leftrightarrow \bar{z} .\label{q1}
\end{align}
The operator picks out difference between boundary values of log $u$ piece of  $A_A$ which is 0. \\ %Rest of the terms vanish due to same reason as we can check as follows :\\\\
\textbf{Term $h^2_{rr}\ \partial_u{F}_{uA}^0$}
\begin{align}
Q_2^{\text{cor}}&=-\frac{1}{2}\int du\ d^2z\  V^A  \partial_u\ \Big[u^2 \partial_u[h^2_{rr}\ \partial_u\overset{0}{F}_{uA}]\Big].
\end{align}
$h_{rr}^2$ has atmost a $\mathcal{O}(u)$ term. Using the $u$-behaviour of $\overset{0}{F}_{uA}$ we see that this term is also 0.\\
\textbf{Term $\gamma^{ BC}h^{-1}_{BA}\ {F}^0_{uC}$}\\
\begin{align}
Q_3^{\text{cor}}&=\frac{1}{2} \int du\  d^2z\  V^A  \partial_u\ \Big[u^2 \partial_u[\gamma^{ BC}h^{-1}_{BA}\ {F}^0_{uC}]\Big].\nn
\end{align}
This term vanishes trivially for classical fall offs of ${F}^0_{uC}$. For the quantum $\log u$ fall offs we get 2 terms for A=z (the analysis is similar for $A$= $\bar{z}$) : 
\begin{align}
Q_3^{\text{cor}}&=-\frac{1}{2} \int du\  d^2z\  V^z  \gamma^{ \bar{z}z}\ \partial_uh^{-1}_{zz}\ A^{0,\log}_{\bar{z}}+\frac{1}{2} \int du\  d^2z\  V^z  \gamma^{ \bar{z}z}\ \partial_uA^0_{\bar{z}}\  h^{-1,\log}_{zz}\ .
% &=\frac{1}{2} \int   d^2z\  V^{z}\ \big[ ... \big]
\end{align}
Upto unimportant overall factors that are common to both terms, the first integrand is : $\lim_{\omega\rightarrow 0}\omega[c_{+}(\omega)+c^\dagger_{-}(\omega)]\   \lim_{\omega\rightarrow 0}\omega[a_{-}(\omega)-a^\dagger_{+}(\omega)]$. Similarly the second integrand is : $\lim_{\omega\rightarrow 0}\omega[c_{+}(\omega)-c^\dagger_{-}(\omega)]\  \lim_{\omega\rightarrow 0}\omega[a_{-}(\omega)+a^\dagger_{+}(\omega)].$ Thus, $Q_3^{\text{cor}}=0.$\\
\textbf{Term $h^0_{rC}\ \partial_u{F}_{AB}^0$}
\begin{align}
Q_4^{\text{cor}}&=\frac{1}{2}\int du\ d^2z\  V^A  \partial_u\ \Big[u^2 \partial_u[ \gamma^{CB}\ h^0_{Cr}\ \partial_u\overset{0}{F}_{AB}]\Big],\nn\\
&=\frac{1}{2}\int du\ d^2z\   \gamma^{CB}\ h^0_{Cr}\ V^A  \partial_u\ [u^2 \partial^2_u\partial_{(B}A^0_{A)}].
\end{align}
This is similar to \eqref{q1} and vanishes by same logic. The analysis for rest of the terms is exactly similar.

\section{Herdegen like representation for graviton}\label{C}

The usual momentum space expression for free metric field is :
\begin{align}
{h_{\mu\nu}}(x)\ &=\sum_{r=+,-} \frac{1}{2(2\pi)^3}\int_{0}^\infty \omega d\omega\ d^2q \  \big[ e^{-i\omega(u+r-\hat{q}.\vec{x})}c^{ r}_{\mu\nu}(\omega,q)\  -e^{-i\omega(u+r-\hat{q}.\vec{x})}c^{\dagger r}_{\mu\nu} (-\omega,q)\ \big].\label{mom}
\end{align}
The angular integral can be performed using stationary phase approximation at large $r$, we can obtain following well known expressions \cite{g}:
\be
h_{zz}(u,q) = \frac{r }{2\pi} \int _{-\infty}^\infty d\omega\ e^{-i\omega u }\  \tilde{C}_{zz}(\omega,q) .\
\ee  
where : 
\be \tilde{C}_{zz}(\omega,q) \ =\  \frac{c_{+}(\omega,q) }{2\pi i(1+|z|^2)^2}....\ \omega>0, \ \ \ \ \tilde{C}_{zz}(\omega,q)=  \frac{-\ c^\dagger_- (-\omega,q) }{2 \pi i(1+|z|^2)^2}\ ....\ \omega<0. \label{Czz}\ee
%where $\tilde{C}_{zz}(\omega)$ is defined by :
%
%and 
And 
\be
h_{\bar{z}\bar{z}}(u,q) = \frac{r }{2\pi} \int _{-\infty}^\infty d\omega\ e^{-i\omega u }\  \tilde{C}_{\bar{z}\bar{z}}(\omega,q) .\
\ee  
\be \tilde{C}_{\bar{z}\bar{z}}(\omega) \ =\  \frac{c_{-}(\omega,q) }{2\pi i(1+|z|^2)^2}....\ \omega>0, \ \ \ \ \tilde{C}_{\bar{z}\bar{z}}(\omega)=  \frac{-\ c^\dagger_+ (-\omega,q) }{2 \pi i(1+|z|^2)^2}\ ....\ \omega<0.\ee\label{Czbar}
%In above expression we can write appropriate combination of creation, annihilation operators as inverse Fourier transform of $C_{zz}$ and $C_{z\bar{z}}$ respectively
Thus \eqref{mom} can be rewritten as  :
\begin{align}
{h_{\mu\nu}}(x)  &= \frac{i}{2(2\pi)^2}\int_{-\infty}^\infty \omega d\omega\ d^2q \ (1+|z|^2)^2\ [\varepsilon_{\mu\nu}^-\ \tilde{C}_{zz}+\varepsilon_{\mu\nu}^+\ \tilde{C}_{\bar{z}\bar{z}} ] e^{-i\omega(u+r-\hat{q}.\vec{x})},\nn\\
 &= \frac{i}{2(2\pi)^2}\int_{-\infty}^\infty \omega d\omega\ d^2q \ (1+|z|^2)^2\ \varepsilon_{\mu\nu}^-\ [\int _{-\infty}^\infty du'\ e^{i\omega u' }C_{zz}] e^{-i\omega(u+r-\hat{q}.\vec{x})}\nn\\
  &+ \frac{i}{2(2\pi)^2}\int_{-\infty}^\infty \omega d\omega\ d^2q \ (1+|z|^2)^2\ \varepsilon_{\mu\nu}^+\  [\int _{-\infty}^\infty du'\ e^{i\omega u' }C_{\bar{z}\bar{z}}] e^{-i\omega(u+r-\hat{q}.\vec{x})},\nn\\
   &=- \frac{1}{(4\pi)}\int d^2q \ (1+|z|^2)^2\ \Big[ \varepsilon_\mu^-\varepsilon_\nu^-\ \dot{C}_{zz}(u=-x\cdot q,\hat{q}) +  \varepsilon_\mu^+\varepsilon_\nu^+\ \dot{C}_{\bar{z}\bar{z}}(u=-x\cdot q,\hat{q})\ \Big],
\end{align}
here, $C_{AB}=\lim_{r\rightarrow \infty}\frac{1}{r}h_{AB}$.
Above expression is analogous to the expression for gauge field obtained by Herdegen \cite{herdegen}.

%\section{References}

\end{document}